\documentclass[extra, mreferee]{gji_joint} 
\usepackage{timet}

\usepackage[pdftex]{graphicx}
\graphicspath{{./Fig/LowRes/}}

\usepackage{setspace}
\usepackage{setspace}
\usepackage{texments}
\usepackage{amssymb}
\usepackage{float}
\usepackage[fleqn]{amsmath}
\usepackage{amsfonts}
\usepackage{algorithm}
\usepackage{algorithmicx}
\usepackage{url}

\let\nl\undefined
\usepackage[linesnumbered, algoruled, lined, commentsnumbered, algo2e]{algorithm2e}
\usepackage{algpseudocode}

\title[Multi-physics PGI]
  {Petrophysically and geologically guided multi-physics inversion using a dynamic Gaussian mixture model}
\author[Thibaut Astic, Lindsey J. Heagy and Douglas W. Oldenburg]
  {Thibaut Astic$^1$, Lindsey J. Heagy$^2$, and Douglas W. Oldenburg$^1$\\
  $^1$ Geophysical Inversion Facility, Department of Earth, Ocean and Atmospheric Sciences,\\University of British Columbia, Vancouver, BC, Canada\\
  $^2$Department of Statistics, University of California Berkeley, Berkeley, CA, United States
  }
\date{August 2020}
\pagerange{}
\volume{}
\pubyear{}

\begin{document}

\label{firstpage}

\maketitle

\begin{summary}
In a previous paper, we introduced a framework for carrying out petrophysically and geologically guided geophysical inversions. In that framework, petrophysical and geological information is modelled with a Gaussian Mixture Model. In the inversion, the Gaussian Mixture model serves as a prior for the geophysical model. The formulation and applications were confined to problems in which a single physical property model was sought, and a single geophysical dataset was available. In this paper, we extend that framework to jointly invert multiple geophysical datasets that depend on multiple physical properties. The petrophysical and geological information is used to couple geophysical surveys that, otherwise, rely on independent physics. This requires advancements in two areas. First, an extension from a univariate to a multivariate analysis of the petrophysical data, and their inclusion within the inverse problem, is necessary. Second, we address the practical issues of simultaneously inverting data from multiple surveys and finding a solution that acceptably reproduces each one, along with the petrophysical and geological information. To illustrate the efficacy of our approach and the advantages of carrying out multi-physics inversions coupled with petrophysical and geological information, we invert synthetic gravity and magnetic data associated with a kimberlite deposit. The kimberlite pipe contains two distinct facies embedded in a host rock. Inverting the datasets individually, even with petrophysical information, leads to a binary geological model: background or undetermined kimberlite. A multi-physics inversion, with petrophysical information, differentiates between the two main kimberlite facies of the pipe. Through this example, we also highlight the capabilities of our framework to work with interpretive geologic assumptions when minimal quantitative information is available. In those cases, the dynamic updates of the Gaussian Mixture Model allow us to perform multi-physics inversions by learning a petrophysical model.

\end{summary}


\section{Introduction}

Mineral deposits, or other geologic features, are characterized by different physical properties, and hence multiple geophysical surveys can be used to delineate them. For example, kimberlites have signatures that depend upon density, magnetic susceptibility, and electrical conductivity. They are often discovered through data collected in airborne surveys and appear as circular low gravity anomalies, with high magnetic responses, and sometimes negative electromagnetic transient responses \citep{Macnae, Keating, Bournas}. Although a joint interpretation of several individually inverted datasets can significantly improve our understanding of the subsurface \citep{OldenburgMilligan1997, TKCpaper, TKCIP, PostInversionClustering0,  PostInversionClustering1, PostInversionClustering3, PostInversionClustering2,  Melo2017}, multiple cases studies have shown that multi-physics inversions can reveal information that was not accessible through individual geophysical dataset inversions \citep{Doetsch2010, Jegen2009, Kamm2015, Lelievre2016}. An extensive compilation of integrated imaging methods and their applications can be found in \citet{integratedImaging}. Multi-physics inversions require a coupling term that mathematically describes a relationship between the different physical property models responsible for the geophysical data. Coupling methods generally use one or a combination of structural or physical property relationships.

The first frameworks for joint inversion focused on linking geophysical models through their structural similarities. \citet{Haber1997} defined the structure of a model in terms of the absolute value of its spatial curvature and compared different models to see if variations occurred at the same locations. This idea was further developed by \citet{Gallardo2003} with the introduction of the concept of cross-gradient between geophysical models. This approach has become commonly used, and both \citet{Gallardo2011} and \citet{Meju2016} provide in-depth reviews of the method and its application. However, this strategy has several limitations: 1) \citet{Meju2016} points out that ``not all physical property distributions in the subsurface will be structurally coincident''; 2) it is unable to reproduce documented or expected petrophysical information \citep{Sun2017}. These drawbacks can be overcome by using other coupling methods.

The second coupling approach uses physical property relationships to link geophysical models. Some of the earliest works used empirical constitutive formulae as their physical properties constraint \citep{Afnimar2002, Hoversten2006, Chen2007}. \citet{Stefano2011} combined this approach with the above mentioned cross-gradient method for sub-salt imaging. \citet{Moorkamp2011} compared the constitutive relationship and the cross-gradient approaches on a 3-D synthetic example combining magnetotelluric, gravity, and seismic data. They concluded that, overall, a cross-gradient approach was preferable compared to using constitutive equations because deviations from the constitutive relations resulted in artefacts in the inverted models; in those situations, the cross-gradient method gave consistent satisfactory results. They also pointed out that the cross-gradient method relies on fewer assumptions about the models than the constitutive equations. Some stochastic frameworks have also been proposed that leverage geostatistical tools to define relationships between physical properties. \citet{Chen2012} used a similar coupling approach as in \citet{Bosch2004} \citep{doi:10.1002/9781118929063.ch9} by building a rock-physics model from borehole data to jointly invert seismic and controlled-source electromagnetic data. \citet{Shamsipour2012} used the geostatistical techniques of cokriging and conditional simulation to jointly invert gravity and magnetic data assuming that the auto- and cross-covariances of the density and magnetic susceptibility follow a linear model of coregionalization. On the deterministic side, of which the framework we present belongs, recent frameworks use clustering techniques such as the fuzzy C-means (FCM) algorithm, which was first used in \citet{Paasche2007} and further expanded in \citet{Lelievre2012}. This approach adds a clustering term to the objective function, which allows more flexible relationships between physical properties. Beyond the addition of the FCM term to the objective function, \citet{Sun2015} introduced an iterative update to the cluster centres throughout an inversion for a single physical property, a technique they called guided FCM; this introduced the concept of uncertainties for physical properties into the inversion. In \citet{Sun2016}, they generalized this work to consider multiple physical properties, and further in \citet{Sun2017}, they added tools to their approach to consider various types of correlations between physical properties (linear, quadratic, etc.). \citet{Giraud2017} represented the petrophysical information as a fixed GMM and focused on reducing uncertainties in stochastic geological modelling by linking potential field inversions and geological models through petrophysical information. To this end, they added, to the Tikhonov objective function, a sum of least-squares differences between the GMM function, evaluated at the current model, and reference values representing the likelihood of their prior knowledge. In \citet{Giraud2019}, they modified their formulation of this coupling term to work with a least-squares difference between the log-likelihood of the GMM and their reference values. In both formulations, they required extensive and fixed quantitative petrophysical and geological information. In practice, that information is not often available and might be only qualitative.

In \citet{ggz389}, we presented a petrophysically and geologically guided inversion (PGI) framework that generalized concepts presented in previously published researches. We showed how geological and petrophysical knowledge represented as a univariate GMM could be incorporated in a voxel-based geophysical inversion through a single smallness term. Each contrasting geological unit was represented by a univariate Gaussian distribution, which summarized its physical property signature. Geological information was included in the GMM through its proportions in a manner similar to that of \citet{Giraud2017}. The log-likelihood of the GMM was then used to regularize the geophysical inversion; this is analogous to the approach taken by \citet{Grana2010} and \citet{Grana2017}. Incorporating both petrophysical and geological information into a single smallness term in the regularization had several advantages. First, it did not require adding a term in the classic Tikhonov formulation. Second, this approach brought the petrophysical data to the same level as the geophysical data, which allowed us to define a misfit, with a target value, between the geophysical model and the petrophysical and geological data. The iteration steps were decomposed into a suite of cyclic optimization problems across the geophysical, petrophysical, and geological data. The petrophysical step formalized the idea of learning the physical property mean values described in \citet{Sun2015} and generalized it to enable the variances and proportions of the GMM to be learned as well during the inversion. These updates to the GMM parameters allowed us to work with partial petrophysical information. The geological step built at each iteration a "quasi-geology model" \citep{QuasiGeologicalModel}, based on the current geophysical and petrophysical models. We applied the PGI approach on synthetic and field data, but the analysis was restricted to single datasets and a single physical property.

This study extends the PGI framework to perform multi-physics inversions, involving several physical properties. We show how geological and petrophysical information represented as a multivariate GMM can be used to couple multiple voxel-based geophysical inversions through a single smallness term in the regularization. The updates to the means, covariances and proportions of the GMM are extended to inversions with multiple physical properties, which expands the work of \citet{Sun2016}. Tools for handling various types of relationships between physical properties are designed; this further develops ideas presented in \citet{Sun2017}. These capabilities, and the gains that they generate, are demonstrated on inversions of synthetic gravity and magnetic data.

In this paper, we start by reviewing the key concepts of the PGI approach and generalize them to the case of multiple physical properties and governing equations for performing multi-physics inversions. We then delineate our strategy to address the numerical challenges of finding a solution to the inverse problem that fits each dataset and, at the same time, adequately fits the petrophysical data. Finally, we demonstrate the advantages of performing joint inversions, with various levels of prior knowledge, by using a synthetic model of the DO-$27$ Tli Kwi Cho kimberlite pipe, Northwest Territories, Canada \citep{JansenEtAl2004}.

\begin{figure}
    \includegraphics[width=\columnwidth]{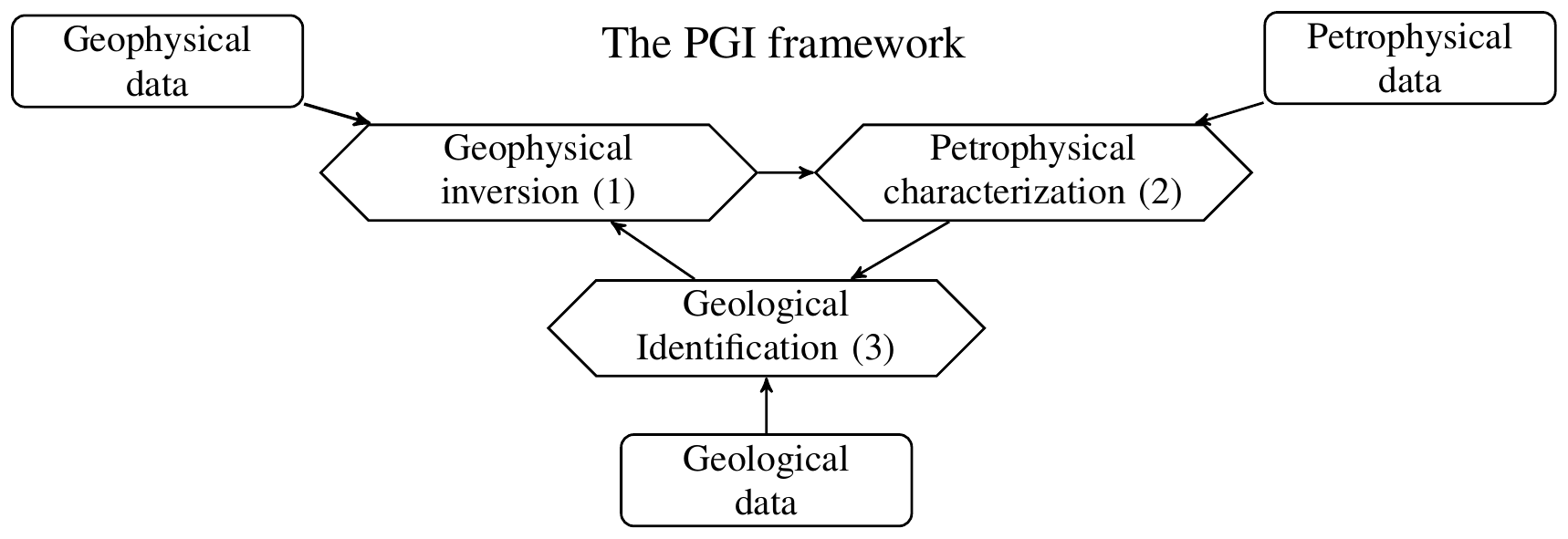}
    \caption{A graphical representation of the PGI framework, modified from \citet{ggz389}. Each diamond box is an optimization process that takes data (shown in rectangular boxes) as well as information provided by the other processes.}
    \label{fig:Framework_with_numbers.pdf}
\end{figure}

\section{Background and motivation for a multi-physics PGI framework}

In this section, we present the conventions we use for notation, provide background on the univariate PGI framework, and motivate its extension to multiple physical properties.

\subsection{Notation conventions}

In terms of notation for parameters, we use the following conventions:

\begin{itemize}
\item Lowercase italic symbols are used for scalar values, such as the trade-off parameter $\beta$.
\item Bold lowercase symbols designate vectors, such as the geophysical model $\mathbf{m}$.
\item Bold uppercase symbols designate matrices, such as the weight matrix $\mathbf{W}$.
\end{itemize}

For a multi-physics inversion, the geophysical model is likely to contain multiple physical properties. Several surveys might be associated with the same physical property (e.g. gravity and gravity gradiometry) or one survey might depend on several physical properties (e.g. electromagnetic surveys depend on both electrical conductivity and magnetic permeability). We thus adopt the following notations for the geophysical model $\mathbf{m}$ indices:
\begin{align}
\mathbf{m} &= vec(\mathbf{M}), \\
\text{with } \mathbf{M} &=
 \begin{pmatrix}
 m_{1, 1} & m_{1, 2} & \cdots & m_{1, q} \\
 m_{2, 1} & m_{2, 2} & \cdots & m_{2, q} \\
 \vdots & \vdots & \ddots & \vdots \\
 m_{n, 1} & m_{n, 2} & \cdots & m_{n, q}
 \end{pmatrix}
\label{m_convention_1}.
\end{align}

A row of the $\mathbf{M}$ matrix represents all the $q$ physical properties that live in the same location. A column represents a single physical property at all the $n$ cells of the mesh. For clarity, we are consistent throughout this study with the index notation. The index $i$ always refers to the cell number, from $1$ to $n$. The vector $\mathbf{m}_i$ then denotes all of the physical properties at the $i$\textsuperscript{th} cell:
\begin{align}
\mathbf{m}_i &= (m_{i, 1}, m_{i, 2}, \dotsc, m_{i, q})^\top.
\label{m_convention_2}
\end{align}

Likewise, we denote the vector model for a single physical property on the whole mesh with the index $p \in \{1..q\}$ with a superscript:
\begin{align}
\mathbf{m}^p &= (m_{1, p}, m_{2, p}, \dotsc, m_{n, p})^\top.
\label{m_convention_3}
\end{align}

Lastly, the geophysical model at iteration $t$ of an inversion is denoted with parentheses $\mathbf{m}^{(t)}$.

\subsection{The Tikhonov inverse problem and its PGI augmentation} \label{section:tik}

The geophysical inverse problem can be posed as an optimization process where the goal is to find a geophysical model $\mathbf{m}$ that minimizes an objective function $\Phi$. Using the same formulation of the inverse problem as in \citet{Tutorial}, the geophysical optimization problem takes the form:
\begin{align}
\begin{split}
&\mathop{\hbox{minimize}}\limits_{\mathbf{m}}\quad\Phi(\mathbf{m}) = \Phi_d(\mathbf{m}) + \beta \left( \alpha_s \Phi_s(\mathbf{m}) + \sum_{v\in{\{x, y, z\}}} \alpha_{v} \Phi_{v}(\mathbf{m}) \right) \label{eq:tikhonov}, \\
&\text{such that}\quad \Phi_d(\mathbf{m}) \leq \Phi_d^*.
\end{split}
\end{align}

In equation \eqref{eq:tikhonov}, the vector $\mathbf{m}$ is the geophysical model, which represents physical properties on a mesh. The term $\Phi_d$ is the geophysical data misfit. The regularization is composed of a smallness term $\Phi_s$, that penalizes model-values different from a reference model, and smoothness terms $\Phi_v$, which penalize variations between adjacent cells; those terms are weighted by positive scaling parameters $\left\{\alpha\right\}$. The trade-off parameter $\beta$ is a positive scalar that adjusts the relative weighting between the regularization and the data misfit. A value of $\beta$ is sought so that the data misfit $\Phi_d$ is below an acceptable target misfit $\Phi_d^*$ \citep{Parker}. The scaling parameters $\alpha_s$ and $\{\alpha_{v}\}$ weight the relative importance of the smallness and smoothness terms. In the Tikhonov inversion, first introduced in \citet{tikhonov1977solutions}, each term of the objective function takes a least-squares form. In particular, the smallness term, which is essential in the PGI framework, can be written as:
\begin{equation}
\Phi_{s}(\mathbf{m})= \frac{1}{2}||\mathbf{W}_{s}(\mathbf{m}-\mathbf{m}_{\text{ref}})||^2_2 \label{regularizer},
\end{equation}
where $\mathbf{m}_{\text{ref}}$ is a reference model and the matrix $\mathbf{W}_s$ represent local weights.

The PGI approach can be considered as an augmentation of the well-established Tikhonov inversion. This is detailed in \citet{ggz389}, where we start from a probabilistic formulation of the least-squares inverse problem \citep{Tarantola}, to then include petrophysical and geological information in the form of a GMM. The resulting term is analogous to the smallness, but the reference model $\mathbf{m}_{\text{ref}}$ and the weights $\mathbf{W}_{s}$ are updated at each iteration. The minimization of the geophysical objective function is labelled Process $1$ in the PGI framework (Fig. \ref{fig:Framework_with_numbers.pdf}).

\subsection{Multivariate GMM: modelling multiple physical properties} \label{PetrophysicsSection}

To extend the approach presented in \citet{ggz389}, we represent the petrophysical signature of each geological unit $j$ ($j=1..c$) as a multivariate Gaussian probability distribution, denoted by $\mathcal{N}$.
The Gaussian function representing the probability distribution of the $q$ physical properties of interest for each unit is defined by its mean $\mitbf{\mu}_j$ (vector of size $q$), and its covariance matrix $\mitbf{\Sigma}_j$ (matrix of size $q\times q$), plus its proportion $\pi_j$.

The multivariate Gaussian Mixture Model (GMM) simply sums the Gaussian probability distribution representing each known unit, weighted by their proportion:
\begin{equation}
\mathcal{P}(\mathbf{x}|\Theta) = \sum_{j=1}^c \pi_j \mathcal{N}(\mathbf{x}_i|\mitbf{\mu}_j, \mitbf{\Sigma}_j) \label{GMM},
\end{equation}
where the variable $\Theta$ holds the GMM global variables $\Theta= \left\{\pi_j, \mitbf{\mu}_j, \mitbf{\Sigma}_j\right\}_{j=1..c}$. With some modifications, the GMM can also represent nonlinear relationships, such as polynomial as presented in \citet{Onizawa2002}. We present those modifications in Appendix \ref{LinearWithMappingSection}, along with an example of an inversion with various nonlinear relationships between two physical properties.
\newpage
The geological classification (or membership) is denoted $\mathbf{z}$. It is defined as the most probable geological unit, given a set of values $\mathbf{x}$ for the $q$ physical properties:
\begin{equation}
z = \mathop{\hbox{argmax}}\limits_{j \in \left\{1..c\right\}} \pi_{j} \mathcal{N}(\mathbf{x}|\mitbf{\mu}_j,\mitbf{\Sigma}_j) \label{eq:MAP_z_physProp}.
\end{equation}

The categorical variable is key for building a "quasi-geology model" \citep{QuasiGeologicalModel} from the physical properties model obtained by inversion. This corresponds to Process $3$ in the PGI framework (Fig. \ref{fig:Framework_with_numbers.pdf}).

\begin{figure}
    \centering
    \includegraphics[width=\columnwidth]{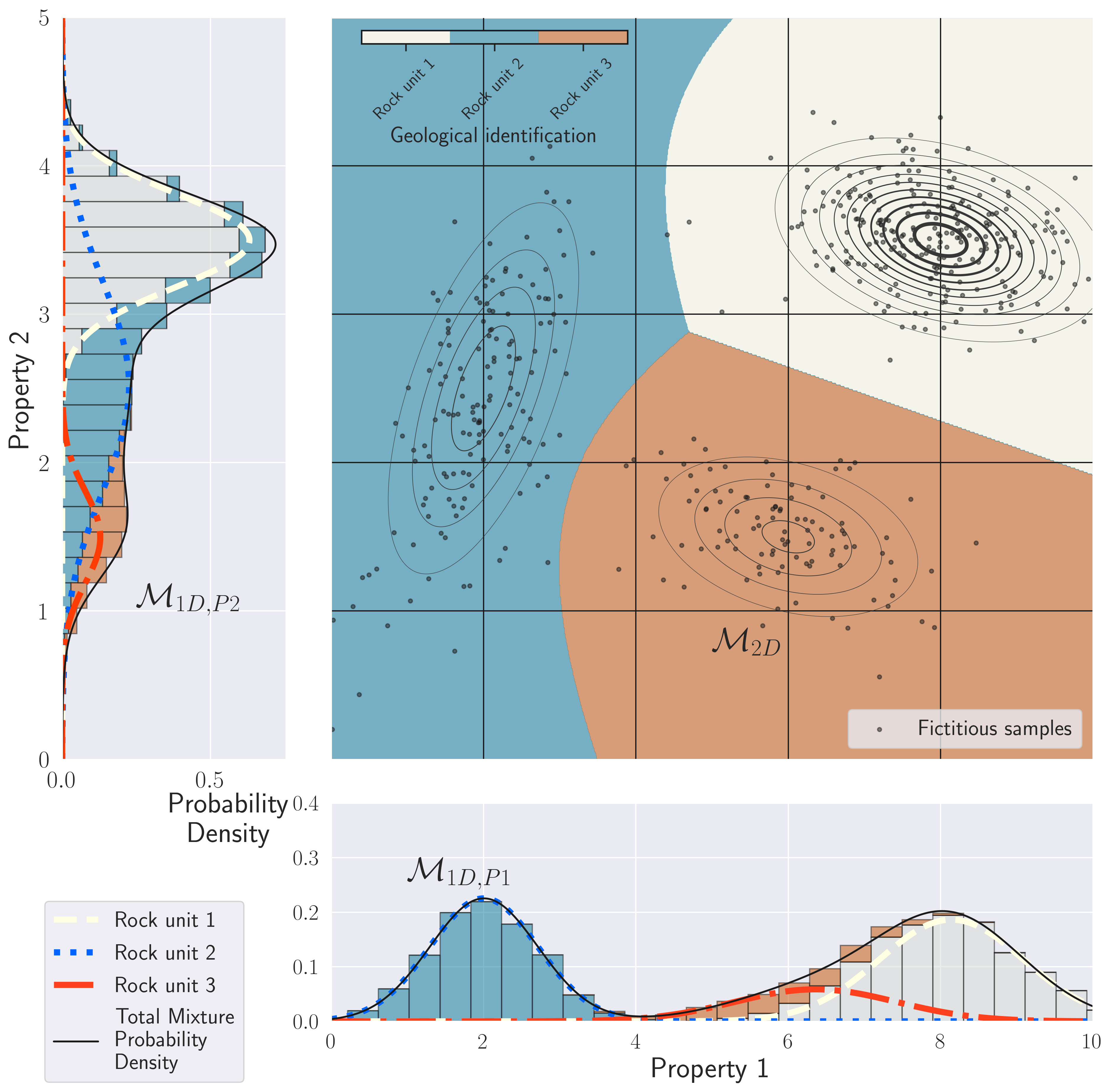}
    \caption{Example of a two-dimensional GMM with three rock units. The background is coloured according to the geological classification evaluated by equation \eqref{eq:MAP_z_physProp}. A thicker contour line indicates a higher iso-probability density level. On the left and bottom panels, we provide the 1D projections of the total and individual probability distributions for each physical property, and the cumulative histograms of the fictitious samples of each rock unit.}
    \label{fig:GaussianMixture_3C_1and2D.png}
\end{figure}

\subsection{Motivation for simultaneously inverting multiple physical properties}

In Fig. \ref{fig:GaussianMixture_3C_1and2D.png}, we present an example of a GMM with three distinct rock units characterized by two physical properties (two-dimensional distributions). The background is coloured according to the geological identification that would be made at each location using equation \eqref{eq:MAP_z_physProp}. The bottom and left panels represent the marginal GMM probability distribution for each physical property individually. We notice that, while all three units are distinct in the two-dimensional space, they can overlap significantly when only considering one physical property at the time. For physical property $1$, rock units $1$ and $2$ are distinct while rock unit $3$ is indistinguishable from rock unit $1$. For physical property $2$, rock units $1$ and $3$ are now distinct while unit $2$ is indistinguishable from either rock units $1$ or $3$. This highlights that it is only by jointly inverting for several physical properties that we might be able to uniquely identify three rock units. Units that might not be distinguishable in one survey may be in another one, and by simultaneously working with both physical properties in an inversion, we are able to explore the $2$ (or multi)-dimensional physical property space in the centre panel of Fig. \ref{fig:GaussianMixture_3C_1and2D.png}. In the following section, we show how to use this probability distribution that links the various physical properties as \textit{a priori} information to regularize the multi-physics inverse problem.

\section{Extension of the PGI framework to multi-physics inversions}

\subsection{Definition of the GMM prior} \label{section:GMMprior}

\citet{ggz389} developed a GMM smallness prior to include petrophysical and geological information in inversions involving only one type of geophysical survey with only one physical property to recover. In equation \eqref{eq:mixturemodel}, we propose a generalized version of the GMM smallness prior that is designed to couple multiple physical properties and incorporate geological information. Note that now the means are vectors that are the size of the number of different physical properties, and the scalar variance becomes a full positive-definite matrix of the same size. The parameters are both spatially (index $i$) and lithologically (index $j$) dependent.
\begin{equation}
\mathcal{M}(\mathbf{m}|\Theta) = \prod_{i=1}^n \sum_{j=1}^c \mathcal{P}(z_{i}=j)\mathcal{N}(\mathbf{m}_i|\mitbf{\mu}_j, \mathbf{W}_i^{-\top}\mitbf{\Sigma}_j\mathbf{W}_i^{-1}),
\label{eq:mixturemodel}
\end{equation}
where:
\newpage
\begin{itemize}
\item $c$ is the number of distinct rock units.
\item $n$ is the number of active cells in the mesh.
\item $\mathbf{m}_i$ represents the physical property values at location $i$.
\item $\mathcal{P}(z_{i}=j)$ is the \textit{a priori} probability of observing rock unit $j$ at location $i$. It can be either constant over the whole area, then denoted by $\pi_j$, or locally determined by \textit{a priori} geological knowledge.
\item $\mitbf{\mu}_j$ contains the means of the physical properties of rock unit $j$.
\item $\mitbf{\Sigma}_j$ is the covariance matrix of the physical properties of rock unit $j$.
\item $\mathbf{W}_i^{-1}$ is a weighting term at location $i$, used for example to include depth or sensitivity weighting. We define it from weights $\left\{{w}_{i, p}, ~i=1..n, ~p=1..q\right\}$; this is a scalar value for each cell $i$ and physical property $p$. $\mathbf{W}_i$ is defined as a diagonal matrix made of the combination of all the weights at the specific location $i$: $\mathbf{W}_i = \text{diag}(\mathbf{w}_{i})$, with the same notation convention as for the model $\mathbf{m}$. This allows the weighting to be different for each physical property. We can thus weight each physical property according to the survey on which it depends.
\item $\Theta$ holds the GMM global variables $\Theta= \left\{\pi_j, \mitbf{\mu}_j, \mitbf{\Sigma}_j\right\}_{j=1..c}$.
\end{itemize}

This GMM probability distribution (equation \eqref{eq:mixturemodel}), representing the current geological and petrophysical knowledge, is used to define the smallness term in the regularization. In the next subsection, we use this multivariate GMM prior to develop a modified objective function for the inverse problem.

\subsection{The multi-physics PGI geophysical objective function}

In \citet{ggz389}, we demonstrated how to use the negative log-likelihood of a univariate GMM as the smallness term in the Tikhonov inverse problem. The resulting smallness term could be approximated by a least-squares misfit between the current model and a reference model $\mathbf{m}_{\text{ref}}$, which was updated at each iteration, as was the smallness matrix $\mathbf{W}_s$. Those dynamic reference model and smallness matrix updates were determined based on the current geophysical model and the geological and petrophysical prior information. The goal of the least-squares approximation was to enable the use of the PGI framework with compiled codes working with the Tikhonov formulation.

Generalizing the result obtained in \citet{ggz389} to multiple physical properties, we use the negative log-likelihood of the GMM defined in equation \eqref{eq:mixturemodel} to obtain a single smallness term that couples all of the model parameters. That smallness term can be approximated by the following least-squares misfit:

\begin{align}
&\Phi_s(\mathbf{m}) = \frac{1}{2}\sum_{i=1}^n||\mathbf{W}_{s}(\Theta, z_i)(\mathbf{m}_i-\mathbf{m}_{\text{ref}}(\Theta, z_i))||_2^2 \label{eq:smallness_petro},\\
&\text{with:}\nonumber\\
&z_i = \mathop{\hbox{argmax}}\limits_{\tilde{z}_i}\mathcal{P}(\mathbf{m}|\tilde{z}_i)\mathcal{P}(\tilde{z}_i) \label{eq:membership}, \\
&\mathbf{m}_{\text{ref}}(\Theta, z_i) = \mitbf{\mu}_{z_i} \label{eq:mref_update},\\
&\mathbf{W}_{s}(\Theta, z_i) = \mitbf{\Sigma}_{z_i}^{-1/2} \mathbf{W}_i\label{eq:Ws_update},
\end{align}
where $\mitbf{\Sigma}^{-1/2}$ is the upper triangular matrix from the Cholesky decomposition of the precision matrix $\mitbf{\Sigma}^{-1}$.

Note that our implementation can handle either the log-likelihood of the GMM or its least-squares approximation. When the petrophysical signature of the rock units are not known, it is possible to learn the parameters of the GMM \citep{ggz389} (Fig. \ref{fig:Framework_with_numbers.pdf}, Process $2$). The extension of that learning process to multiple physical properties can be found in Appendix \ref{UpdateTheta}.

\subsection{Petrophysical target misfit} \label{sec:target}

The PGI smallness expresses a misfit between the petrophysical and geological information and the geophysical model. To measure the goodness of fit and define a stopping criterion for the petrophysical misfit, \citet{ggz389} defined a measure $\Phi_{\text{petro}}$ and its target value $\Phi_{\text{petro}}^*$. This measure is similar to the PGI smallness term but without the weights $\mathbf{W}_i$ in equation \eqref{eq:Ws_update}. The same approach can be taken here to define the value $\Phi_{\text{petro}}$ for the multivariate case:
\begin{align}
&\Phi_{\text{petro}}(\mathbf{m}) = \frac{1}{2}\sum_{i=1}^n||\mathbf{\tilde{W}}_{s}(\Theta, z_i)(\mathbf{m}_i-\mathbf{m}_{\text{ref}}(\Theta, z_i))||_2^2 \label{eq:petroness_petro},\\
&\text{with: } \nonumber \\
&\mathbf{\tilde{W}}_{s}(\Theta, z_i) = \mitbf{\Sigma}_{z_i}^{-1/2},
\end{align}
where $z_i$ and $\mathbf{m}_{\text{ref}}$ are the same as in equations \eqref{eq:membership} and \eqref{eq:mref_update}.

Looking at the term $\Phi_{\text{petro}}$ (equation \eqref{eq:petroness_petro}) from a probabilistic point of view \citep{Tarantola,ggz389}, each variable $\mathbf{m}_i$ follows a multivariate Gaussian of dimension $q$, with mean $\mathbf{m}_{\text{ref}}(\Theta, z_i)$, and covariance matrix $\left(\mathbf{\tilde{W}}_{s}(\Theta, z_i)^\top \mathbf{\tilde{W}}_{s}(\Theta, z_i)\right)^{-1}$. Thus, the variable $\mathbf{\tilde{W}}_{s_i}(\Theta, z_i)(\mathbf{m}_i-\mathbf{m}_{\text{ref}_i})$ follows a multivariate Gaussian variable with mean $\mathbf{0}$ and an identity covariance matrix. Thus, the sum in $\Phi_{\text{petro}}$ follows a chi-squared distribution and we can apply Pearson's chi-squared test \citep{Pearson1900}. The target misfit value $\Phi_{\text{petro}}^*$ is defined as the expectation of $\Phi_{\text{petro}}$:
\begin{equation}
 \Phi_{\text{petro}}^* = E[\Phi_{\text{petro}}] = \frac{n\cdot q}{2} \label{eq:petrotarget},
\end{equation}
with $n$ being the number of active cells in the mesh and $q$ being the number of physical properties. This generalizes the result obtained in \citet{ggz389} ($q=1$). This is a similar approach to the definition of a target misfit for the geophysical data as given in \citet{Parker}.

Our algorithm stops when all of the target misfits, geophysical and petrophysical, are achieved. In the next section, we present our strategy for handling multiple geophysical data misfits as well as an additional petrophysical misfit, each with its target value we seek to reach.

\section{Numerical considerations for reaching multiple target misfits}\label{section:Implementation}

We have multiple geophysical data misfits that we wish to fit. The inclusion of petrophysical and geological data with PGI adds another data misfit term that also needs to reach its target misfit (section \ref{sec:target}). In this section, we provide our strategies for choosing and dynamically adjusting the various parameters of the objective function to find a solution that fits all the data. An algorithm that summarizes the whole framework is provided in Appendix \ref{algorithm}.

\subsection{Objective function with multiple geophysical data misfits}

The objective function we seek to minimize for the multi-physics inversion process (Fig. \ref{fig:Framework_with_numbers.pdf}, Process $1$) takes the form:
\begin{align}
&\Phi(\mathbf{m}) = \Phi_d(\mathbf{m}) + \beta \left( \alpha_s \Phi_s(\mathbf{m}) + \sum_{v\in{\{x, y, z\}}}\sum_{p=1}^q\alpha_{v, p}{\Phi_{v, p}(\mathbf{m})}\right) \label{fullobjfct},\\
&\text{with: } \nonumber \\
&\Phi_d(\mathbf{m}) = \sum_{k=1}^r \chi_k \Phi_d^k(\mathbf{m}) = \frac{1}{2} \sum_{k=1}^r \chi_k ||\mathbf{W}_{d}^k(\mathbb{F}^k\lbrack\mathbf{m}^{\{k\}}\rbrack-\mathbf{d}_{\text{obs}}^k)||^2_2, \label{eq:datamisfit} \\
&\Phi_{v, p}(\mathbf{m}) = \frac{1}{2}||\mathbf{W}_{v,p}\mathbf{L}_v(\mathbf{m}^p-\mathbf{m}^p_{\text{ref}})||_2^2 \label{eq:smoothnessip}.
\end{align}

The data misfit term $\Phi_d(\mathbf{m})$ now contains multiple geophysical data misfits, each defined as a weighted least-squares norm. $\Phi_d^k$ is the data misfit of the $k$\textsuperscript{th} survey, where the forward operator $\mathbb{F}^k$ generates the predicted data for that survey from $\mathbf{m}^{\{k\}}$. The notation $\mathbf{m}^{\{k\}}$ denotes the subset of model parameters associated with the $k$\textsuperscript{th} survey. Note that multiple surveys can be associated with the same physical property, for example, gravity and gravity gradiometry both depend on density contrasts. Similarly, one survey can be sensitive to several physical properties; for example, electromagnetic surveys are sensitive to electrical conductivity and magnetic susceptibility. The data measured by the $k$\textsuperscript{th} survey is symbolized by $\mathbf{d}_{\text{obs}}^k$ and the uncertainty on those measurements by the matrix $\mathbf{W}_{d}^k$. Each data misfit $\Phi_d^k$ is weighted by a scaling parameter $\chi_k$. Those $\{\chi\}$ scaling parameters are important for balancing the various geophysical data misfits and finding a solution that fits all of them. Our approach for updating these parameters is developed later in this section.

The regularization is still composed of the smallness and smoothness terms. The smallness term $\Phi_s$ is our coupling term, which is defined in equation \eqref{eq:smallness_petro}. The smoothness terms, one for each direction and physical property, are represented by $\Phi_{v, p}$. In the smoothness terms (equation \eqref{eq:smoothnessip}), the smoothness operators (usually first or second-order difference) are represented by the matrix $\mathbf{L}_v$, and weights (sensitivity or depth) are represented by the matrix $\mathbf{W}_{v,p}$.

Equation \ref{fullobjfct} is an intricate objective function that is the sum of many quadratic regularization terms, each of which is multiplied by an adjustable constant. Finding values for these constants and carrying out a nonlinear inversion to produce a model that acceptably fits the data, and is a good candidate for representing information from the true geology model, is numerically challenging.

In the following sections, we present our approach for estimating and updating these parameters throughout the inversion. The smoothness scaling parameters $\{\alpha_{v, p}\}$ are the only values that we keep constant. The scaling parameter $\alpha_s$ weights the importance of the petrophysical misfit term, and the trade-off parameter $\beta$ influences the importance of the regularization (which contains the petrophysical misfit) relative to the geophysical data misfits. The geophysical data misfit scaling parameters $\{\chi\}$ are used to adjust the relative importance of each geophysical dataset. Values of $\beta$, $\alpha_s$, and $\{\chi\}$ are sought so that each geophysical data misfit $\Phi_d^k$ is below or equal to its target misfit ${\Phi_d^k}^*$, along with a value of the petrophysical data misfit $\Phi_{\text{petro}}$ that is less than or equal to its target misfit $\Phi_{\text{petro}}^*$.

\subsection{The regularization scaling parameters $\beta$ and $\{\alpha\}$}

Two types of scaling parameters act on the regularization terms; they are the trade-off parameter $\beta$ and the $\{\alpha\}$ parameters. In our implementation, we keep the $\{\alpha_{v,p}\}$ parameters acting on the smoothness terms constant while we update $\beta$ and $\alpha_s$ to reach a suitable solution to the PGI problem. Next, we outline our strategies for each of these parameters.

\subsubsection{Fixed parameters: the smoothness scaling parameters $\{\alpha_{v, p}\}$}

In the smoothness terms, the scaling parameters $\{\alpha_{v, p}\}$ control the relative importance of spatial derivative terms in the regularization. Each set of assigned values will yield different outcomes. This is often a way in which model space can be explored (e.g. preferential smoothness in some directions, see \citet{Williams_2008, Lelievre2009}). They are generally specified \textit{a priori}, and we keep them fixed in our objective function throughout the inversion process. In addition to the common practical considerations for choosing the smoothness parameters \citep{Tutorial,williams2006applying}, we use the $\{\alpha_{v, p}\}$ to weight each physical property, by dividing it by the square of its expected maximum amplitude (available through the GMM means if provided). This helps equalize the contribution of the smoothness terms to the objective function value when parameters have widely different scales (like density, log- electrical conductivity and magnetic susceptibility contrasts). The scaling of the physical properties in our extended smallness term (equation \eqref{eq:smallness_petro}) is taken care of by the covariance matrices of the GMM.

\subsubsection{Adjusted parameters: $\beta$ and $\alpha_s$}

In the case of a single geophysical data misfit with a petrophysical misfit, \citet{ggz389} developed a strategy for cooling $\beta$ and warming $\alpha_s$ to find a solution to the inverse problem that reaches the target values of both misfits. This approach is still appropriate in the multi-physics inversion framework and is what we use in this study (step 6 in algorithm \ref{algo:algorithm}). For the multi-physics case, we alter it in the following way: when all geophysical data misfits are equal or below their target value, the strategy for warming the scaling parameter $\alpha_s$ on the coupling term is:
\begin{equation}
\alpha_s^{(t+1)} = \alpha_s^{(t)} \cdot \mathop{\hbox{median}}\limits_{k=1..r} \left(\frac{{\Phi_d^{k}}^*}{{\Phi_d^{k}}^{(t)}}\right) \label{eq:alpha_warm}.
\end{equation}


\subsection{Balancing the geophysical data misfits with the scaling parameters $\{\chi\}$}


Our goal is to develop a strategy for scaling multiple geophysical data misfits so that each geophysical dataset is adequately fit. We propose a strategy where the scaling parameters $\left\{\chi_k\right\}_{(k=1..r)}$ in equation \eqref{eq:datamisfit} are successively updated. Our approach has a heuristic foundation and does not incur the significant computational cost often associated with optimization-based approaches, and generalize to any number of geophysical data misfits. Before presenting its details, we first outline some strategies that others have taken in addressing this problem.

\subsubsection{Review of previous strategies for balancing various geophysical data misfits and coupling terms}

Several approaches for weighting multiple geophysical data misfits have been proposed in the recent literature. Some frameworks do not follow any prescribed strategy for updating the scaling parameters of the geophysical data misfits. This is the case for the approaches proposed by \citet{Sun2016, Sun2017} and \citet{Sosa}; both keep those scaling parameters constant. \citeauthor{Sun2017} use values of unity, while \citeauthor{Sosa} normalize each geophysical data misfit by its number of data. In our experience, keeping the weights constant has led us to overfit some surveys while underfitting others. Other frameworks have adopted the approach of running their joint inversions for multiple combinations of parameters. For three geophysical data misfits, \citet{Moorkamp2011} adopted a manual check-and-guess approach to adjust the parameters. For two geophysical data misfits, \citet{Giraud2019} ran a subset of their inversion hundreds of times for various combinations of scaling parameters before choosing values based on the L-curve principle \citep{HansenLcurve0, Hansenlcurve, SantosLcurve}. They then manually "fine-tuned" those values using the full joint inversion problem. To avoid the issue of having to choose multiple appropriate scaling parameters, \citet{Bijani2017} developed a compromise between deterministic and stochastic optimizations for joint inversions. They adopted a "Pareto Multi-Objective Global Optimization" strategy with genetic algorithms that generate populations of candidate models that "simultaneously minimize multiple objectives in a Pareto-optimal sense," rather than working with a fully aggregated objective function. This approach was still computationally expensive and limited to small 2-D studies in the paper.

To limit the number of multiple runs of the same inversion, \citet{Lelievre2012} devised a rigorously defined, but computationally expensive, strategy for dynamically balancing two geophysical data misfits. Their approach relied on adjusting first the trade-off parameter until the two data misfits are Pareto-optimal. Next, it adjusted the relative weights of the two surveys to fit both geophysical surveys. It then reinforced the importance of the coupling term before going into another round of adjustments of the trade-off and surveys weights parameters. The approach developed in \citet{ggz389} for a single geophysical data misfit, but with a petrophysical data misfit, is related to the work of \citet{Lelievre2012}. Both focused first on fitting the geophysical data misfit terms and then adjusting the coupling term. On the contrary, the strategy presented in \citet{gallardo_strategy} favoured the cross-gradients coupling over the geophysical misfits.

Here we define a practical, computationally inexpensive, heuristic strategy for balancing the geophysical data misfits as well as the coupling term. We design this strategy to work for any number of surveys, and thereby generalize the work of \citet{Lelievre2012}. For the full algorithm, the reader can refer to Appendix \ref{algorithm}.

\subsubsection{Strategy for updating $\{\chi\}$}

We now define our strategy for weighting the multiple geophysical data misfits to reach all the target values. We use the scaling parameters $\left\{\chi_k\right\}_{(k=1..r)}$ defined in equation \eqref{eq:datamisfit}. We dynamically update each geophysical misfit scaling parameter based on its current misfit and target value, compared to the other surveys.

We start with a set of initial scaling parameters $\{\chi\}$ that sums to unity. To ensure the progress of all data misfits, while limiting the possibilities of overfitting any given term, we update the scaling parameters $\{\chi\}$ during the inversion. Our approach is philosophically similar to what is proposed in \citet{ggz389} for $\beta$ and $\alpha_s$ in order to balance the geophysical data misfit and the petrophysical misfit at each iteration in the inversion, and generalizes ideas proposed in \citet{Lelievre2012} to more than two geophysical data misfits. If one geophysical data misfit reaches its target value before the others, we use the ratio of its current value with its target to warm the scaling parameters of the other geophysical data misfit terms. We then normalize the sum of the scaling parameters to be equal to unity again; this is to keep the importance of the total $\Phi_d$ term relatively similar before and after adjusting the scaling parameters $\{\chi\}$. If several surveys are below their respective targets, we simply use the median of the ratios to warm the scaling parameters of the still unfit surveys. Thus, at any iteration $(t)$ of the geophysical inverse problem, if an ensemble of $\left\{k_f\right\}$ surveys has reached their respective targets, we warm the scaling parameters of the remaining $\left\{k_u\right\}$ surveys that are not yet fit as (step $7$ in algorithm \ref{algo:algorithm}):
\begin{align}
&\tilde{\chi}_{k_u}^{(t+1)} = \chi_{k_u}^{(t)} \cdot \mathop{\hbox{median}}\limits_{\left\{k_f\right\}} \left(\frac{{\Phi_d^{k_f}}^{(t)}}{{\Phi_d^{k_f}}^*}\right) \label{eq:chi_update}, \\
&\tilde{\chi}_{k_f}^{(t+1)} = {\chi}_{k_f}^{(t)}, \\
&\text{then we normalize the sum:} \nonumber \\
&{\chi^{(t+1)}_k} = \frac{\tilde{\chi}_k^{t+1}}{\sum_{k=1}^r \tilde{\chi}_{k}^{(t+1)}} \label{eq:chi_normalizing}.
\end{align}

An example of convergence curves for the data misfits and evolution of the dynamic scaling parameters is proposed in Fig. \ref{fig:TKC_ConvergenceCurves.png} for a multi-physics PGI with full petrophysical information.

\begin{figure}
\centering
\includegraphics[width=\columnwidth]{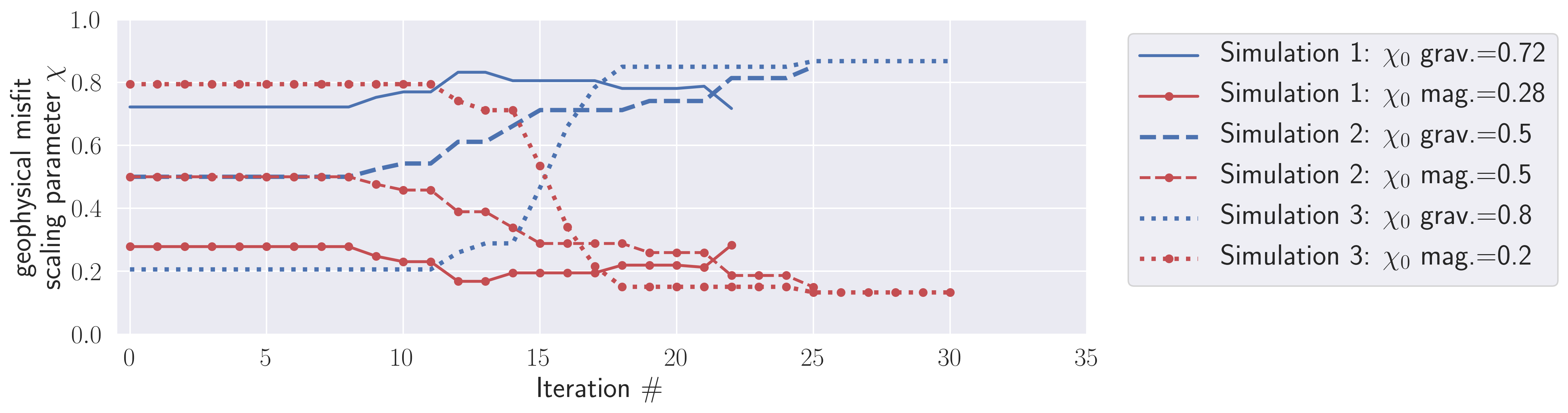}
\caption{Evolution curves of the scaling parameters $\{\chi\}$ with the proposed strategy for three multi-physics PGIs with full petrophysical information and different initializations for $\{\chi\}$. The color of each line corresponds to the geophysical misfit: blue for gravity and red with markers for magnetic. The style of the lines corresponds to one of the three inversions ($\chi_{0,\text{grav}}+\chi_{0,\text{mag}}=1$ in each inversion).}
\label{fig:TKC_ChiCurves.png}
\end{figure}

Our strategy has proven to be insensitive to the initialization of the scaling parameters $\{\chi\}$ for linear problems. To demonstrate this point, we show in Fig. \ref{fig:TKC_ChiCurves.png} the evolution of the scaling parameters $\{\chi\}$ for three multi-physics PGI with full petrophysical information. The synthetic example presented in section \ref{sec:multiphysics_full} is run with various initializations $\{\chi_0\}$. The outcomes of all these three PGI were similar to the result we show in Fig. \ref{fig:TKC_Joint_Synthetic.png}. The scaling parameters $\chi$ associated with the magnetic and gravity misfits, respectively, all finish at approximately the same value even though the initializations are very different. The final $\chi$ scaling values are about $0.8$ for the gravity data misfit and around $0.2$ for the magnetic data misfit. This is an appealing property as it reduces the need to fine-tune the initialization of the scaling parameters $\{\chi\}$, which can be costly for large-scale inversions.

\section{Numerical implementation}

We implemented our framework as part of the open-source software \texttt{SimPEG} \citep{Cockett2015, heagy2017framework}. As such, we are able to share both the software environment and the scripts to reproduce the examples shown in this paper \citep{PGIJointExamples}. In this section, we highlight some key points of our implementation to encourage the use of this work and future collaborations. A more detailed tutorial is provided in Appendix \ref{pseudocode}, which lays out a pseudocode sketch of the implementation.

\texttt{SimPEG} is designed to be a modular, extensible framework for simulations and inversions of geophysical data. In particular, two features enabled us to focus our implementation efforts on the PGI framework, while using tools provided by the open-source community (such as the forward operators):

\begin{enumerate}
\item the composability of objective functions in the data misfit and the regularization terms,
\item the \texttt{directives}, which orchestrate updates to components of the inversion at each iteration.
\end{enumerate}

The first point enables the implementation of joint inversions. In the code, each misfit term is a Python object that has properties, such as the weights used to construct $\mathbf{W}_d$, and methods, including functions to evaluate the misfit given a model as well as derivatives for use in the optimization routines. To construct a gravity and magnetic joint inversion, we first define each misfit term independently and then sum them. We use operator-overloading in Python so that when we express the addition of two objective functions in code, the evaluation of this creates a \texttt{combo-objective function}. This is an object that has the same evaluation and derivative methods as the individual data misfits, and thus readily inter-operates with the rest of the simulation and inversion machinery in \texttt{SimPEG}.

To the second point, \texttt{directives} are functions that are evaluated at the beginning or end of each iteration in the optimization. They are the mechanism we use to update to components of the inversion, including the data misfit scaling parameters (equation \eqref{eq:chi_update}), smallness weights and reference model (equations \eqref{eq:mref_update} and \eqref{eq:Ws_update}), and for evaluating the target misfits and stopping criteria for the inversion.

\section{Example: The DO-$27$ kimberlite pipe} \label{TKC}

In this section, we illustrate the joint PGI approach on synthetic gravity and magnetic data based on the DO-$27$ kimberlite pipe \citep{JansenEtAl2004}, which is composed of two different kimberlite facies. We compare standard Tikhonov inversions of the individual geophysical datasets and independent PGIs of the gravity and magnetic data with the multi-physics PGI approach. Both the Tikhonov inversions and single-physics PGIs produce models that enable only a binary distinction: kimberlite or host rock. Only the multi-physics PGI allows us to identify the two kimberlite facies as distinct from the background host rock.

Scripts and Jupyter notebooks to reproduce the examples presented in this study are available through \texttt{GitHub} at \url{https://github.com/simpeg-research/Astic-2020-JointInversion} \citep{PGIJointExamples}.

\subsection{Setup}

\begin{figure*}
    \centering
    \includegraphics[width=\textwidth]{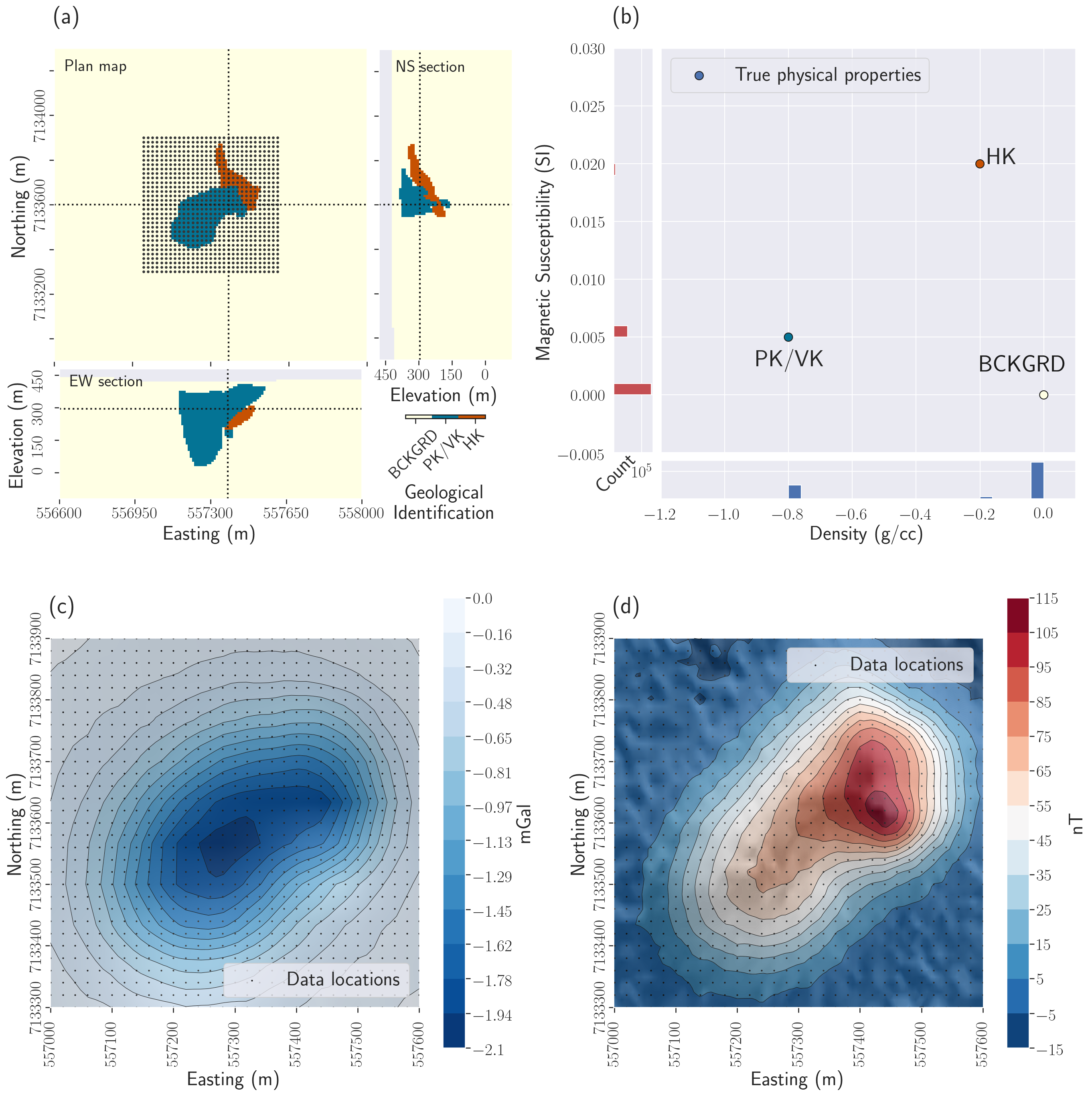}
    \caption{Setup: DO-$27$ synthetic example: (a) Plan map, East-West and North-South cross-sections through the synthetic geological model. The grid of dots represents the data locations for the gravity and magnetic survey; the dotted lines represent the location of each cross-section. (b) Cross-plot and histograms of the physical properties of the synthetic model; (c) Synthetic ground gravity data; (d) Synthetic total amplitude magnetic data.}
    \label{fig:TKC_Setup_Synthetic.png}
\end{figure*}

The DO-$27$ Kimberlite pipe (Northwest Territories, Canada) is part of a complex known as the Tli Kwi Cho (TKC) kimberlite cluster \citep{JansenEtAl2004} (Fig. \ref{fig:TKC_Setup_Synthetic.png}). The pipe has two distinctive kimberlite units that are embedded in a background consisting of a granitic basement covered by a thin layer of till (Fig. \ref{fig:TKC_Setup_Synthetic.png}a). The first pipe unit is a pyroclastic and volcanoclastic kimberlite (called PK/VK), which has a weak magnetic susceptibility and a very high negative density contrast. The second unit is a hypabyssal kimberlite (called HK), which has a strong magnetic susceptibility and a weak negative density contrast. The Tikhonov inversions of the field gravity and magnetic datasets have been documented in \citet{TKCpaper}.

For this example, we use simulated surface gravity and airborne magnetic data modelled from a synthetic model of the DO-$27$ pipe. The forward and inversion mesh is a tensor mesh with 375 442 active cells; each has a pair of density-magnetic susceptibility values. The smallest cells are cubes with a $10$ m edge length. All chosen values for the surveys and geological units are consistent with observations documented in \citet{TKCpaper}. For the PK/VK unit, we assume a magnetic susceptibility of $5\cdot10^{-3}~\text{SI}$ and a density contrast with the background of $-0.8~\text{g/cm}^3$. For the HK unit, the magnetic susceptibility is set to $2\cdot10^{-2}~\text{SI}$ and the density contrast to $-0.2~\text{g/cm}^3$ (Fig. \ref{fig:TKC_Setup_Synthetic.png}b). We forward modelled the data over a grid of 961 receivers, at the surface for the gravity survey and at the height of $20$ m for the airborne magnetic survey (Fig. \ref{fig:TKC_Setup_Synthetic.png}c and d). We added unbiased Gaussian noise to the gravity and magnetic data with standard deviations of $0.01$ mGal and $1$ nT, respectively. These standard deviations are input into the data weighting matrices $\left\{{\mathbf{W}_d^k, ~k=1, 2}\right\}$.
\newpage
For each inversion, we added bound constraints so that the sought density contrast values are null or negative, and the magnetic susceptibility contrast values are null or positive. We used the sensitivity of each survey to define the $\left\{\mathbf{w}_{ip}, ~i=1..n, p=1..q\right\}$ weights. Each physical property is weighted by the sensitivity of its associated survey. Sensitivity-based weighting is a common practice for potential fields inversions \citep{Li1996, Li1998, Portniaguine2002, SensW}. The initial model is the background half-space for all inversions.

\begin{figure*}
    \centering
    \includegraphics[width=\textwidth]{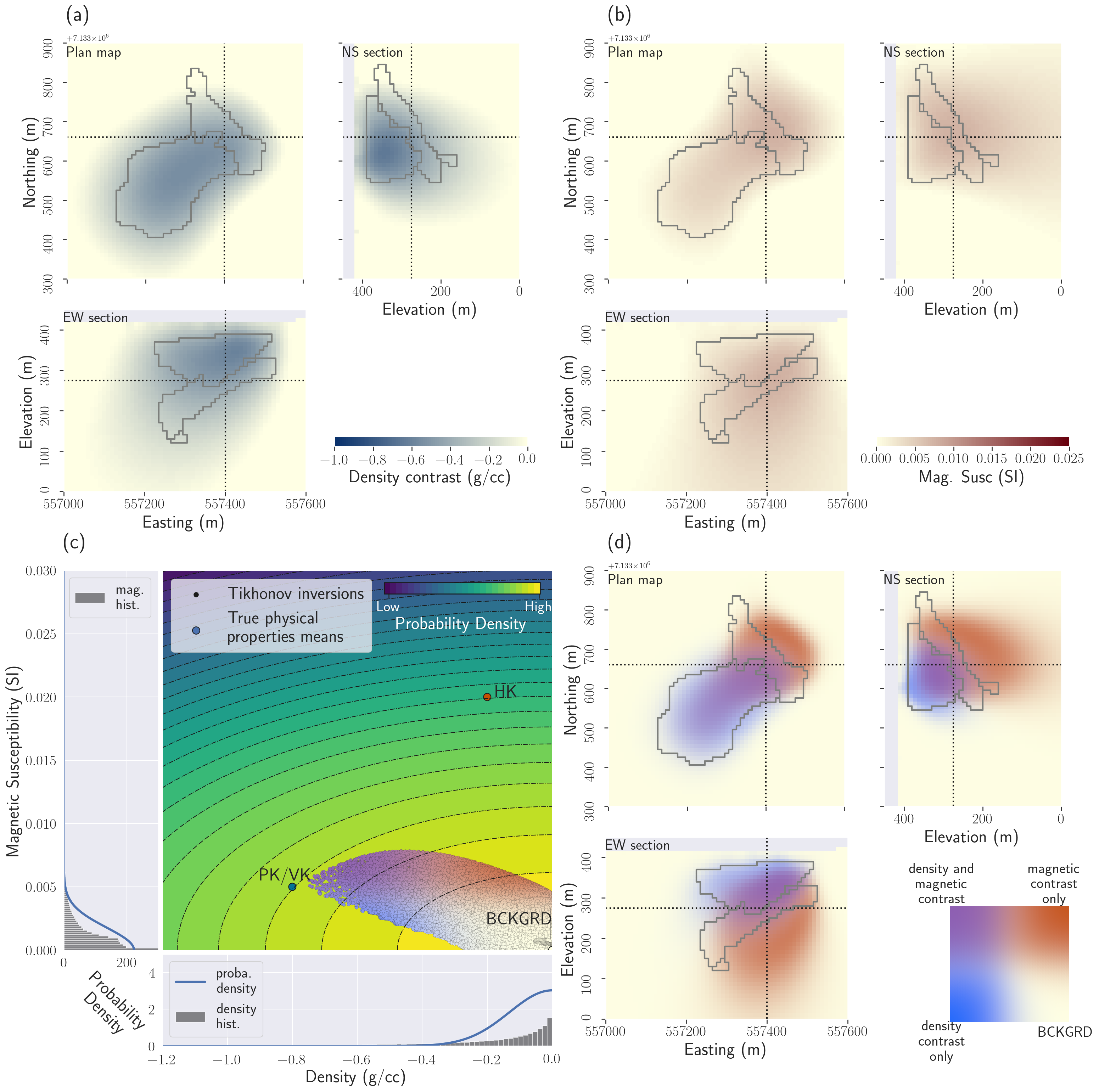}
    \caption{DO-$27$ gravity and magnetic Tikhonov inversion results. (a) Plan map, East-West and North-South cross-sections through the recovered density contrast model; (b) Plan map, East-West and North-South cross-sections through the recovered magnetic susceptibility contrast model; (c) Cross-plot of the density and magnetic susceptibility models. The points are coloured using the density and the magnetic susceptibility contrast values (white for the background (BCKGRD), blue for density contrast only, red for magnetic susceptibility contrast only, and purple for co-located significant density and magnetic susceptibility contrasts). The side and bottom panels show the marginal distribution of each physical property, with the best fitting univariate Gaussian (proba. stands for probability, and hist. stands for histogram). Those two univariate Gaussian distributions are used to compute the multivariate Gaussian showed in the background of the cross-plot; (d) Plan map, East-West and North-South cross-sections coloured based on the combination of density and magnetic susceptibility contrasts recovered by Tikhonov inversions (same colourmap as used in (c)).}
    \label{fig:TKC_L2_Synthetic.png}
\end{figure*}

\subsection{Tikhonov inversions}

We first run the individual inversions of the gravity and magnetic data using the well-established Tikhonov approach described in Section \ref{section:tik}. The results, shown in Fig. \ref{fig:TKC_L2_Synthetic.png}, are relatively smooth. The gravity inversion (Fig. \ref{fig:TKC_L2_Synthetic.png}a) provides an approximate outline of the pipe. The magnetic inversion (Fig. \ref{fig:TKC_L2_Synthetic.png}b) shows a body centred on HK, but it is too diffuse to delineate a shape. Fig. \ref{fig:TKC_L2_Synthetic.png}(c) shows the cross-plot of the recovered density and magnetic susceptibility models. Each point is coloured based on both its density and magnetic susceptibility values: white when both contrasts are low, with a blue-scale for a significant density contrast only, with a red-scale for only a significant magnetic susceptibility, and with a purple-scale when both contrasts are significant. We observe the expected continuous Gaussian-like distribution of the model parameters (in the region allowed by the bound constraints). Petrophysical signatures are not reproduced, with notably the strongest density contrasts being co-located with the highest magnetic susceptibility values (mesh cells coloured in purple in Fig. \ref{fig:TKC_L2_Synthetic.png}c and d); this is in contradiction with the setup where PK/VK, the unit with a low density, is distinct from HK, the unit with high magnetic susceptibility. In both gravity and magnetic inversions, the two anomalous units are indistinguishable from each other. To highlight this, we show an overlap of the two inversions in Fig. \ref{fig:TKC_L2_Synthetic.png}(d). We coloured each point relative to its density and magnetic susceptibility values, as in Fig. \ref{fig:TKC_L2_Synthetic.png}(c). This juxtaposition highlights that combining both models does not show structures that seem closer to the ground truth. Post-inversions classification would give highly variable results, depending on the thresholds chosen to delineate units.

We next move to a PGI approach and include petrophysical information. We start by inverting each geophysical dataset individually, and we assess what gains are made before moving to a multi-physics PGI.

\subsection{Single-physics PGIs}

\begin{figure*}
\centering
\includegraphics[width=\textwidth]{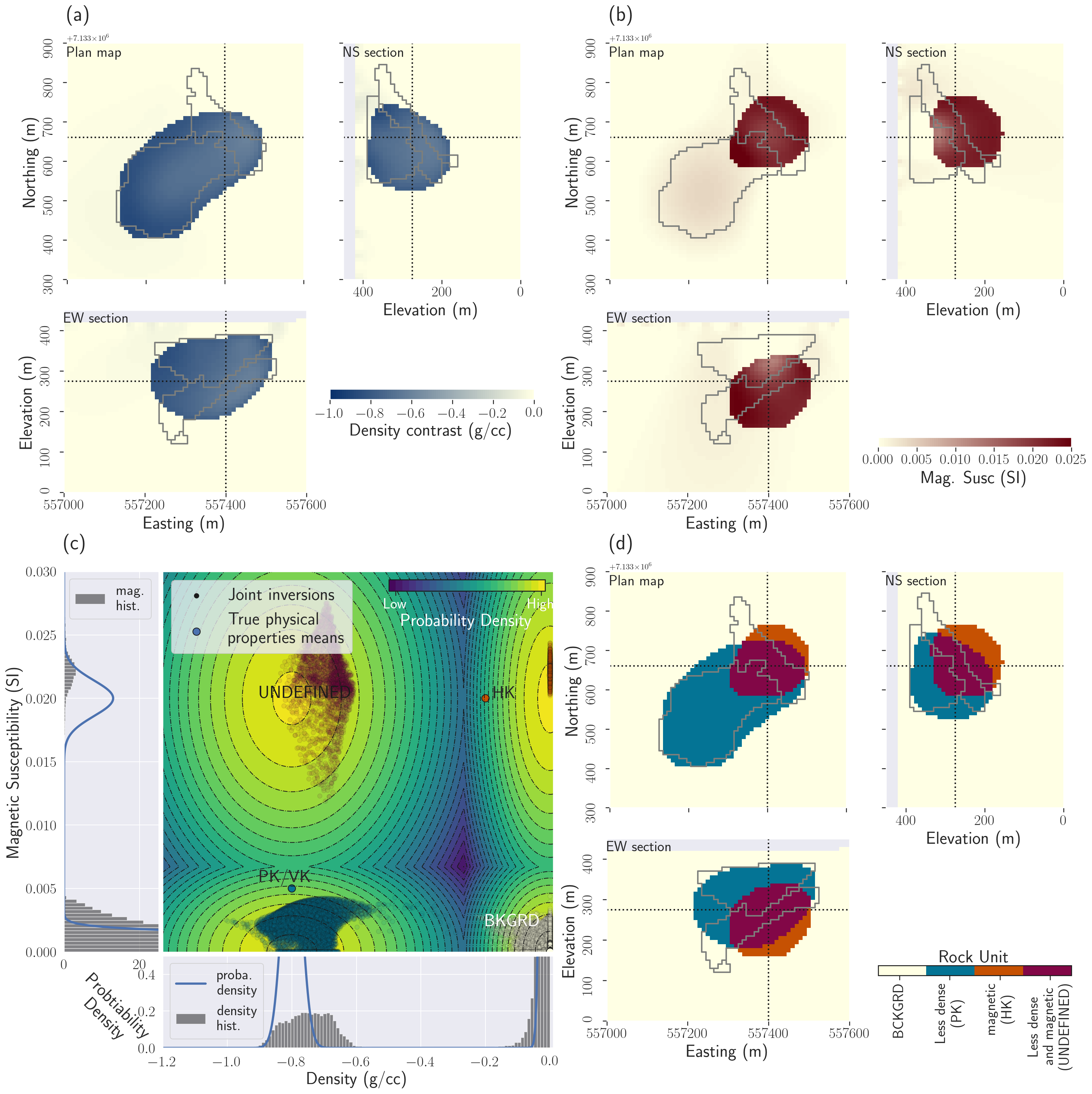}
\caption{Results of the individual PGIs. (a) Plan map, East-West and North-South cross-sections through the density model recovered using the petrophysical signature of PK/VK; (b) Plan map, East-West and North-South cross-sections through the magnetic susceptibility model obtained using the petrophysical signature of HK; (c) Cross-plot of the inverted models. The 2D distribution in the background has been determined by combining the two 1D distributions used for density and magnetic susceptibility PGIs, respectively. With only one anomalous unit in each case, there is still four possible combinations; (d) Plan map, East-West and North-South cross-sections through the quasi-geology model built from the density and magnetic susceptibility models, see cross-plot in (c).}
\label{fig:TKC_IndividualPetro_Synthetic.png}
\end{figure*}

We apply the PGI framework developed in \citet{ggz389} to invert each geophysical dataset individually. The results are shown in Fig. \ref{fig:TKC_IndividualPetro_Synthetic.png}. For the prior petrophysical distribution, we use the true value for the means of each unit. For the petrophysical noise levels, we assign standard deviations of $3.5\%$ of the highest known mean value for each physical property except for the background for which we assign $1.75\%$ (see the 1D left and bottom panels for the distribution of each physical property, respectively, in Fig. \ref{fig:TKC_IndividualPetro_Synthetic.png}c). For the proportions, we also used the true values. We acknowledge that proportion values could be difficult to estimate in practice. However, in our experiments, the values of the global proportions have not had a significant impact on the inversion result. The use of locally varying proportions can, however, guide the reproduction of particular features \citep{Giraud2017, ggz389}. All the GMM parameters are held fixed in those single-physics inversions.

In carrying out the inversions, we found that both magnetic and gravity data can be explained individually by assuming a single unit, either PK/VK or HK. Each dataset, gravity or magnetic, can be fit by either reproducing the signature of PK/VK or HK, or any value in between. The difference in physical property contrast is compensated for by a difference in the volume of the recovered body. The two kimberlite facies are thus indistinguishable when we consider one geophysical survey at the time. Adding a third cluster in either inversion to represent the second kimberlite facies does not help, as it only gives the algorithm more "choices" that are not supported by the data. For conciseness, we choose to show here the gravity result recovered using only the petrophysical signature of the PK/VK unit (Fig. \ref{fig:TKC_IndividualPetro_Synthetic.png}a), which is the most responsible for the gravity anomaly; for the magnetic inversion, we show the model obtained by only using the petrophysical signature of HK, which is the unit that is the most responsible for the magnetic response (Fig. \ref{fig:TKC_IndividualPetro_Synthetic.png}b). The additional models (gravity inversion with HK's density signature and magnetic inversion with PK/VK's magnetic signature) are shown in Appendix \ref{sec:additionalIndividual}; these demonstrate the discrepancy in the recovered volumes of each unit between the magnetic and gravity inversions. For example, explaining the gravity anomaly with only a body with the same density as HK leads to a very large body, bigger than the volume of that same unit recovered through the magnetic inversion. The same reasoning applies to the PK/VK body.

The gravity PGI using the PK/VK unit petrophysical signature (Fig. \ref{fig:TKC_IndividualPetro_Synthetic.png}a) gives useful information about the depth and delineation of the pipe that was not available from the Tikhonov inversion. The magnetic PGI using the HK petrophysical signature (Fig. \ref{fig:TKC_IndividualPetro_Synthetic.png}b) places a body around the HK unit location but misses its elongated shape. From the petrophysical perspective, both the gravity signature of PK/VK and the magnetic signature of HK are individually well reproduced. However, the combination of the density and magnetic susceptibility contrasts recovered by the individual PGIs (cross-plot in Fig. \ref{fig:TKC_IndividualPetro_Synthetic.png}c) is very far from the desired multidimensional petrophysical distributions. Even by assuming just two units for each inversion (background and kimberlite) as we did, there are still four different combinations of density and magnetic susceptibility values. In this specific case, looking at Figs \ref{fig:TKC_IndividualPetro_Synthetic.png}c) and d), there is: 1) a cluster representing the background with both weak density and magnetic susceptibility contrasts (coloured in white); 2) a cluster with a large density contrast and a low susceptibility (coloured in blue); this is close to the petrophysical signature of the PK/VK unit; 3) a cluster with high magnetic susceptibility and very small density contrasts (coloured in orange); This would be the HK unit; 4) a cluster that has both high magnetic susceptibility and large density contrasts, that we identify as 'undefined' in the figures. This last cluster does not correspond to any unit signature and occupies a large volume. This hinders our ability to resolve two clear kimberlite facies from the inversions. Therefore, this motivates us to move to a multi-physics inversion approach to take advantage of the density-magnetic susceptibility relationships in the inversion and finally delineate two distinct kimberlite facies.


\subsection{Multi-physics PGI with petrophysical information} \label{sec:multiphysics_full}

\begin{figure*}
    \centering
    \includegraphics[width=\textwidth]{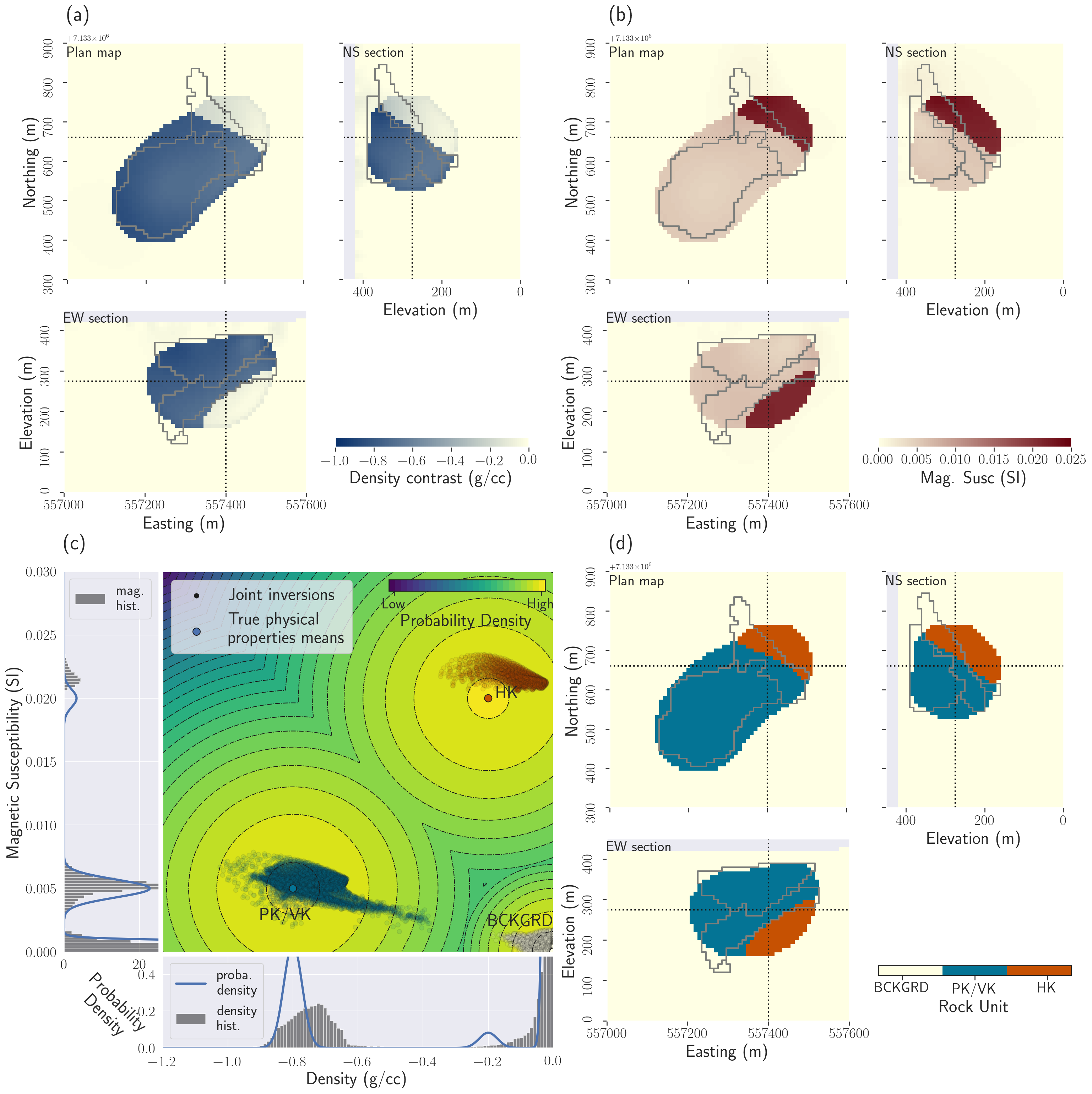}
    \caption{Results of the multi-physics PGI with petrophysical information. (a) Plan map, East-West and North-South cross-sections through the recovered density contrast model; (b) Plan map, East-West and North-South cross-sections through the magnetic susceptibility contrast model; (c) Cross-plot of the inverted models. The colour of the points has been determined by the clustering obtained from this framework joint inversion process. In the background and side panels, we show the prior joint petrophysical distribution with true means we used for this PGI; (d) Plan map, East-West and North-South cross-sections through the resulting quasi-geology model.}
    \label{fig:TKC_Joint_Synthetic.png}
\end{figure*}

We now apply our multi-physics PGI framework to jointly invert the gravity, magnetic data with petrophysical information (means, covariances, proportions for each unit, background, PK/VK, HK, of the GMM). The parameters $\left\{\mathbf{w}_{ij}\right\}_{i=1..n, j=1..c}$ are again used to include the appropriate sensitivity weighting for each method and physical property. We use the same uncertainties that we used for the individual PGIs. The off-diagonal elements of the covariance matrices are set to null, which just means we assume no correlations between the density and magnetic susceptibility variations within a single rock unit. The GMM parameters are still held fixed. The results are shown in Fig. \ref{fig:TKC_Joint_Synthetic.png}.

The improvement is significant. The joint inversion succeeds in recovering two distinct kimberlite facies that reproduce the provided petrophysical signatures. The quasi-geology model (Fig. \ref{fig:TKC_Joint_Synthetic.png}d) is geologically consistent and does not introduce erroneous structures. The surface outline of the pipe is well recovered. The vertical extension is similar to that of the true model. We also now have indications of the elongated shape and tilt of the HK unit. The magnetic susceptibility of HK is slightly overestimated but still within the acceptable margins defined by the petrophysical noise levels we assigned (Fig. \ref{fig:TKC_Joint_Synthetic.png}c).

\begin{figure}
\centering
\includegraphics[width=0.5\textwidth]{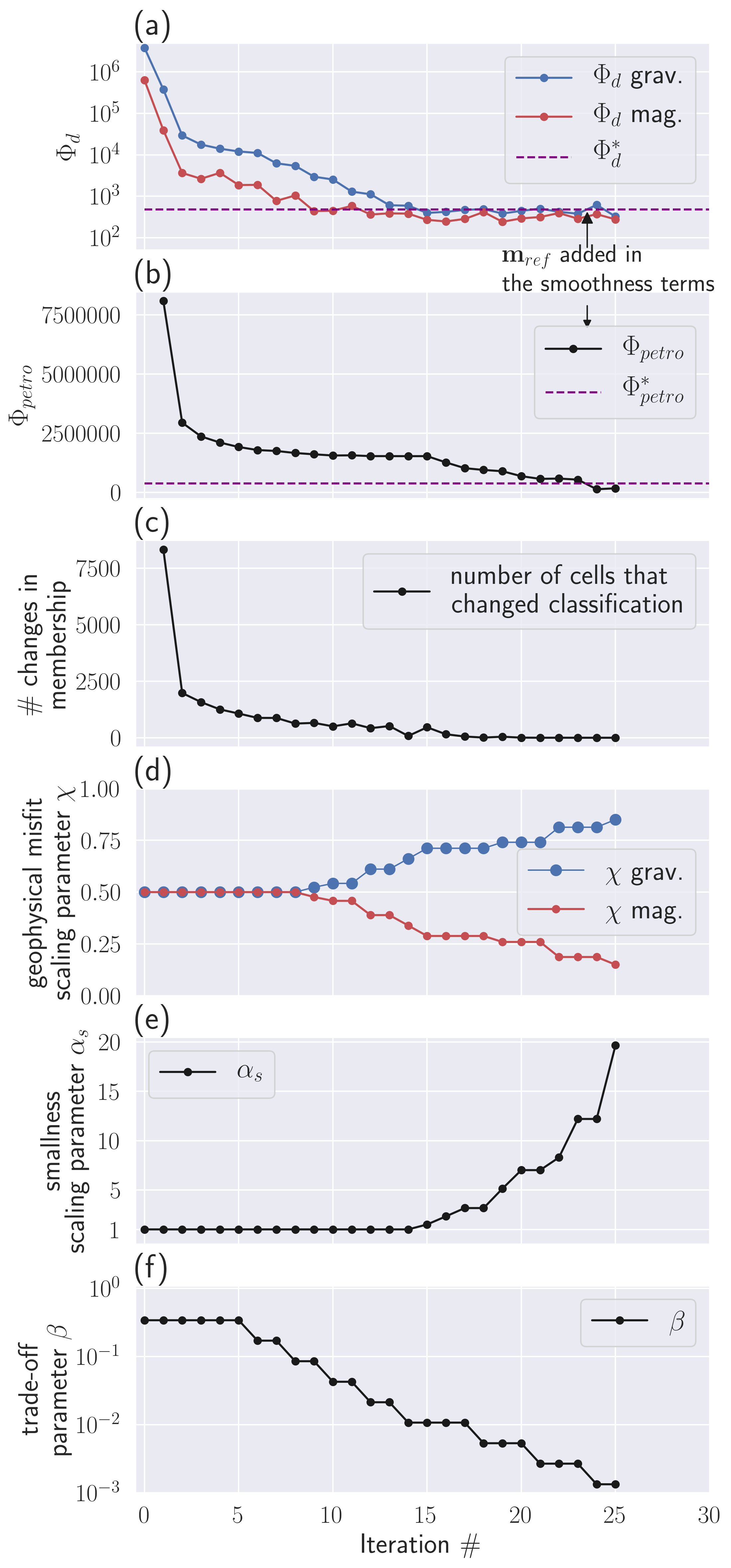}
\caption{Convergence curves for the three misfits, and evolution curves for the dynamic scaling parameters during the multi-physics PGI with petrophysical information shown in Figure \ref{fig:TKC_Joint_Synthetic.png}. (a) Gravity and magnetic geophysical data misfits and their targets (same number of data); (b) Convergence curves of the petrophysical misfit, defined in equation \eqref{eq:petroness_petro}, and its target value, defined in equation \eqref{eq:petrotarget}; (c) Evolution of the $\{\chi\}$ scaling parameters; (d) Evolution of the $\alpha_s$ scaling parameter; (e) Evolution of the trade-off parameter $\beta$.}
\label{fig:TKC_ConvergenceCurves.png}
\end{figure}

To illustrate the behaviour of our heuristic strategy for the update of the objective function scaling parameters, we provide in Fig. \ref{fig:TKC_ConvergenceCurves.png} the convergence curves of the three data misfits (gravity, magnetic, and petrophysical), and the evolution of the various dynamic scaling parameters, for the multi-physics PGI with full petrophysical information. All target values are reached after $25$ iterations, and the PGI stops.

This result was obtained by providing the petrophysical means of the rock units in the GMM. In the next inversion, we devise our approach for using the multi-physics PGI framework when quantitative information is not available.

\subsection{Multi-physics PGI with limited information}

We have illustrated the gains made by the multi-physics PGI framework when extensive and quantitative \textit{a priori} information is provided. We now investigate how to perform multi-physics inversions when \textit{a priori} information to design the coupling term is not available. \citet{ggz389} emphasized the benefits of learning a GMM during the inversion process to compensate for uncertain, or unknown, petrophysical information. At each iteration, the GMM parameters are determined by running a Maximum A Posteriori Expectation-Maximization (MAP-EM) clustering algorithm \citep{ExpectationMaximization}. The MAP-EM algorithm estimates compromise values for the GMM parameters based on the prior GMM parameters, weighted by confidence parameters in this prior knowledge, and the current geophysical model. We generalize the learning process of the GMM parameters to a multidimensional case in Appendix \ref{UpdateTheta}.

In the next example, we demonstrate how learning the means of the kimberlite units iteratively through the inversion allows us to still perform multi-physics inversions without providing physical property mean values. This is done by acting on the $\{\mitbf{\kappa}\}$ confidence parameters in the means. A confidence $\kappa$ value of zero indicates that the mean is fully learned from the inversion, while an infinite confidence fixes the mean to its prior value. In all the inversions with limited information, we fix the means of the background for both physical properties to their true values (zero); this is a usual assumption in Tikhonov inversions of potential fields that the background has a zero contrast. We keep the covariances of the GMM fixed and similar to what we used previously. The covariance matrices define our petrophysical noise levels and how spread each petrophysical signature can be.

\subsubsection{Employing qualitative information with PGI} \label{sec:InterpreterAssumption}

\begin{figure*}
    \centering
    \includegraphics[width=\textwidth]{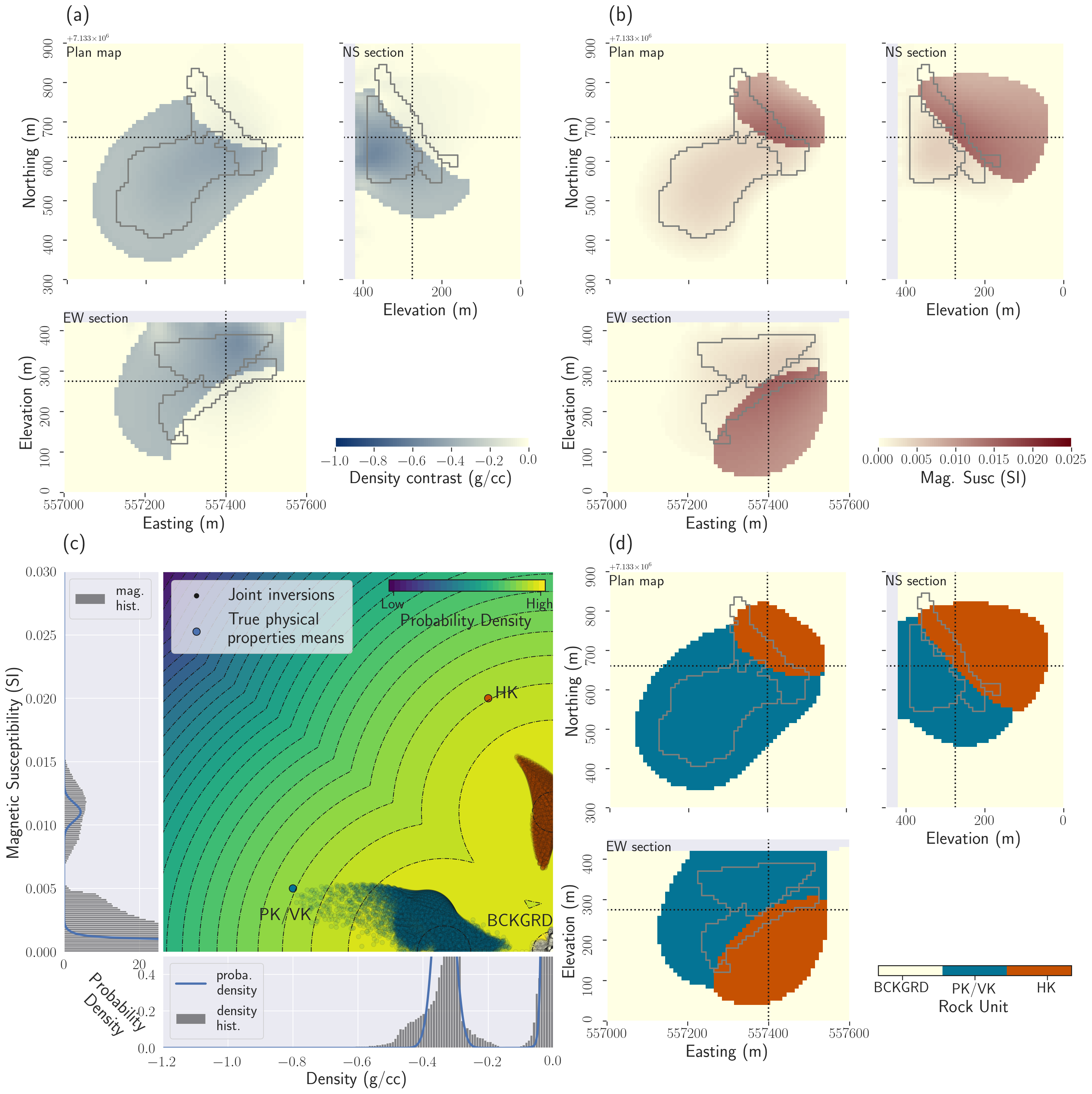}
    \caption{Results of the multi-physics PGI without providing the means of the physical properties for the kimberlite facies, and assuming a low-density unit and a magnetized unit; (a) Plan map, East-West and North-South cross-sections through the recovered density contrast model; (b) Plan map, East-West and North-South cross-sections through the magnetic susceptibility contrast model; (c) Cross-plot of the inverted models. The colour of the points has been determined by the clustering obtained from this framework joint inversion process. In the background and side panels, we show the learned petrophysical GMM distribution; (d) Plan map, East-West and North-South cross-sections through the resulting quasi-geology model.}
    \label{fig:TKC_NoPrior3ClustersAssumption_Synthetic.png}
\end{figure*}

\begin{figure}
\centering
\includegraphics[width=\columnwidth]{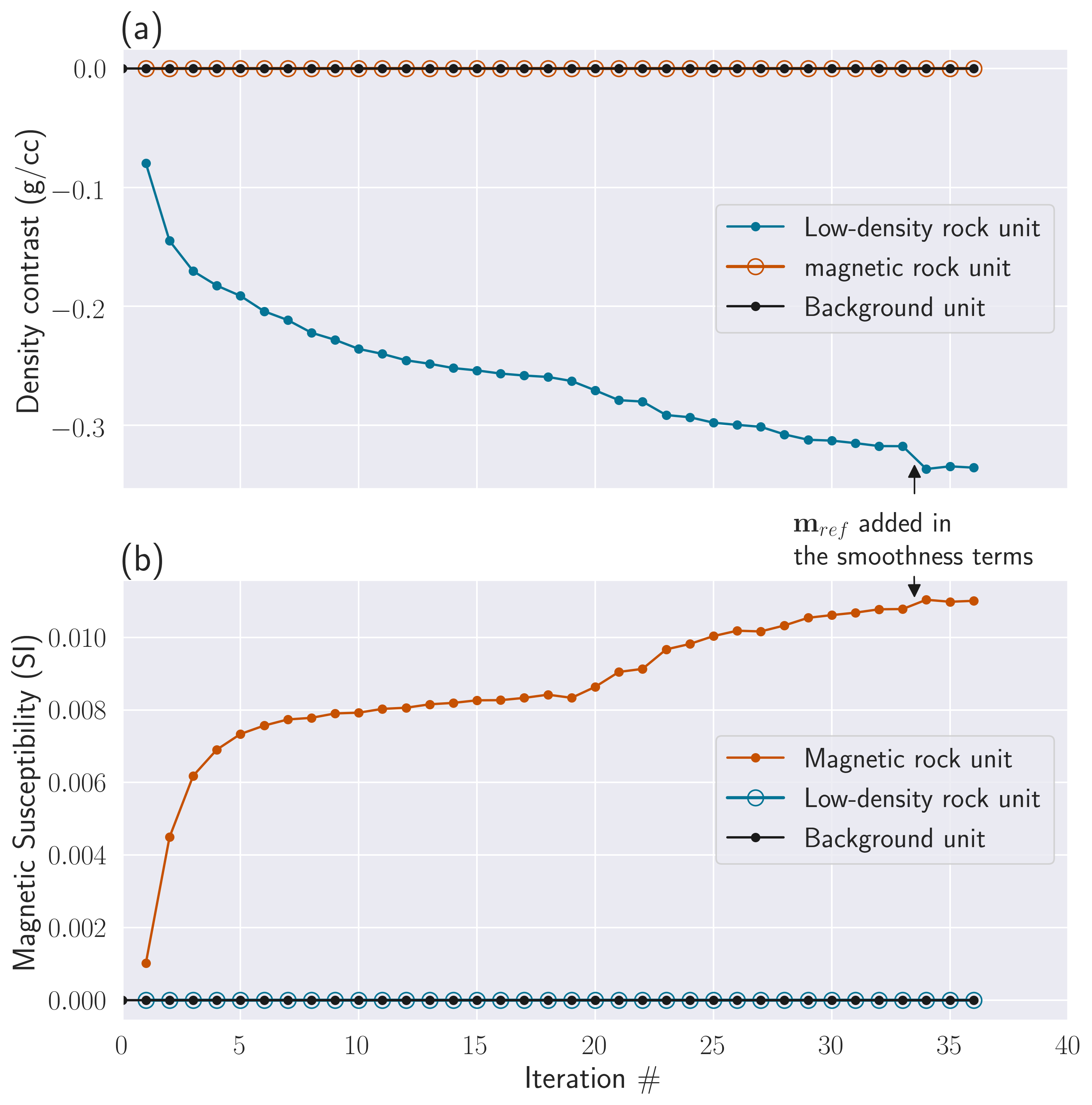}
\caption{Evolution of the learned means of the GMM throughout the multi-physics PGI with qualitative information for the three assumed rock units (background, low-density kimberlite and magnetic kimberlite) shown in Figure \ref{fig:TKC_NoPrior3ClustersAssumption_Synthetic.png}. The background mean values, the density of the magnetic rock unit, and the magnetic susceptibility of the low-density rock unit are kept fixed. Initialization has a low impact on the learned mean values, and thus the values at iteration 0 are not shown in the plot. (a) Evolution of the density contrast mean values ; (b) Evolution of the magnetic susceptibility mean values.}
\label{fig:TKC_LearningGMMmeans.png}
\end{figure}

Once the Tikhonov inversions have been run (see Fig. \ref{fig:TKC_L2_Synthetic.png}), it already appears likely that the gravity and magnetic anomalies are mostly generated by two distinct bodies, as the centres of the two recovered anomalous bodies (density and magnetic susceptibility) are at different locations. Lacking quantitative petrophysical information, the multi-physics PGI framework allows us to formulate the following "interpreter's assumption": one kimberlite unit is responsible for the gravity anomaly, while a second one is responsible for the magnetic response. Employing this assumption is made possible in our framework by defining the confidences in the means of each unit $\{\mitbf{\kappa}\}$ as vectors. This allows us to act on each physical property mean value of each unit. For example, for the kimberlite unit that is assumed to be responsible for the magnetic response, we set the confidence $\kappa$ in its magnetic susceptibility to zero; the MAP-EM algorithm decides its value at each iteration based solely on the current geophysical model. On the contrary, its mean density contrast is kept fixed at zero by setting the confidence $\kappa$ in this mean value to infinity. The same procedure is applied for the kimberlite unit that is assumed to be responsible for the gravity response. The initialization of the density contrast and magnetic susceptibility mean values, for the kimberlite units responsible of the gravity and magnetic response respectively, has little impact on the inversion result, so long as the initial guess is reasonable. It is also common practice, in general, to run clustering algorithms multiple times from various initializations before choosing a specific outcome \citep{ExpectationMaximization, Murphy2012}. In the specific result shown in Fig. \ref{fig:TKC_NoPrior3ClustersAssumption_Synthetic.png}, the density contrast and magnetic susceptibility mean values for the respective kimberlite rock units were initialized at $-1 \text{ g/cm}^3$ and $0.1$ SI. Similar results were obtained with other initializations ($-0.4 \text{ g/cm}^3$ and $0.01$ SI etc., but not with $0 \text{ g/cm}^3$ and $0$ SI for all units). Because of the weak dependency of the result with regard to the initialization, we choose not to show the initial value in Fig. \ref{fig:TKC_LearningGMMmeans.png}, which presents the evolution of the means throughout the multi-physics PGI with qualitative information.


The result of the multi-physics PGI, with no petrophysical information but with the assumption of distinct low-density and magnetized units, is shown in Fig. \ref{fig:TKC_NoPrior3ClustersAssumption_Synthetic.png}. Three distinct clusters (background, low-density unit, magnetized unit) are well recovered. The final learned means are respectively $-0.33 \text{ g/cm}^3$ for the low-density unit and $1.1 \cdot10^{-2}$ SI for the magnetized unit.

This multi-physics inversion has several advantages over any of the single-physics inversions. First, by bringing in a qualitative, geologic assumption, we are able to delineate two units by avoiding the overlap of low density and high susceptibility anomalies. Second, we get a sense of the dip of the HK unit. None of those two achievements was reached by the Tikhonov inversions or the single-physics PGIs, even with petrophysical information.

The means of the GMM are learned iteratively following the constraints defined by our assumptions: the background has a fixed contrast of zero in both physical properties, one rock unit is responsible for the gravity response with a null magnetic contrast, and one rock unit is responsible for the magnetic response with a null density contrast. The evolution of the estimations of the means throughout the inversion is shown in Fig. \ref{fig:TKC_LearningGMMmeans.png}. All target values are reached after $36$ iterations, and the PGI stops.

Finally, we note that distinguishing two bodies, each mostly responsible for a particular geophysical response, is not automatically required from the geophysical datasets themselves. Indeed, we present in Appendix \ref{sec:additionalIndividual}, Fig. \ref{fig:TKC_NoPrior_Synthetic.png} a multi-physics PGI with no petrophysical information and only two clusters, where the confidences $\{\mitbf{\kappa}\}$ are all zeros for the kimberlite unit; the background mean is kept fixed at zero contrast for both physical properties. A single anomalous body is able to fit both gravity and magnetic datasets, with a learned mean and an acceptable spread according to the set covariance matrices. This further highlights the gains made possible by the opportunities to simply incorporate a qualitative, geologic assumption within the inversion framework.

\subsection{DO-$27$ example summary}

From the petrophysical perspective, the density-magnetic susceptibility cross-plot for the standard Tikhonov inversions is very different from the expected distribution (Fig. \ref{fig:TKC_Joint_Synthetic.png}c). The smoothness of the recovered models and physical property distributions does not allow us to delineate and distinguish between the two kimberlite facies clearly. Using PGI, individual datasets can both be reproduced using a single kimberlite facies. The individual gravity PGI gives us more information about the depth and delineation of the main PK/VK body. The individual magnetic PGI yields a reasonable estimate for the depth of the HK unit but misses its elongated shape. The two individually recovered quasi-geology models are, however, incompatible when they are combined because of the significant overlap of the recovered PK/VK and HK units. The multi-physics PGI without petrophysical information produces a geological model that distinguishes between the two kimberlite facies. It also begins to give us information about the elongated shape and dip of the HK unit; this result was not achieved by any of the single-physics inversions, not even by the ones that included petrophysical information. However, the accuracy of the boundaries of the bodies is affected by the lack of petrophysical information. The result is improved by providing petrophysical information to the multi-physics PGI, which yields our best recovered model.

\section{Discussion}

We have expanded the PGI framework developed in \citet{ggz389} to carry out multi-physics joint inversions, and we have proposed a strategy to balance any number of geophysical data misfits along with a coupling term. In our experiments, this strategy appeared to be critical to fitting data from multiple surveys as well as petrophysical data. Finally, we have used a synthetic example to demonstrate the capabilities of the multi-physics PGI framework.

With regards to the iterative learning of GMM means when limited information is available, considering the confidences $\{\mitbf{\kappa}\}$ as vectors is an important contribution of our framework and it advances the approach of \citet{Sun2016}. In their approach, the updates to the means are controlled per unit only, without differentiating the physical properties that are well-known from the undocumented ones. They can either learn the means of a unit or keep it fixed, whereas the framework we present is capable of learning specific components of the GMM means for each unit, as demonstrated in the DO-$27$ example.

We demonstrated examples of multi-physics inversions with potential fields, which are linear problems. In previous works \citep{ggz389}, the PGI approach was applied to nonlinear electromagnetic problems (magnetotelluric, direct-current resistivity, and a field frequency-domain electromagnetic dataset), but it considered only individual surveys depending on a single physical property. We plan to implement this approach for performing multi-physics inversions with electromagnetic methods. This will also be an opportunity to test the robustness of our reweighting strategies for multi-physics inversions with nonlinear geophysical problems. Areas such as the DO-$27$ kimberlite pipe \citep{TKCpaper, TKCEM, TKCIP}, with many different types of geophysical surveys available, are prime candidates for the application of the PGI approach to refine the image of the subsurface structures by integrating more datasets and physical properties in a single inversion.

As we apply the PGI framework to more complex problems, the handling of various types of relationships between physical properties is required. Linear relationships are straight-forward to implement with Gaussian distributions through the covariance matrix, which can define tilted, elongated probability distributions. In Appendix \ref{LinearWithMappingSection}, we discuss how to account for nonlinear relationships. Such nonlinear relationships are found, for example, between density and seismic velocities \citep{Onizawa2002}. Our framework is flexible enough that different relationships can be included for each rock unit. While modest in size, our example in Fig. \ref{fig:LinearWithMapping.png} is, to our best knowledge, the first one in the literature with such diverse relationships in a single inversion. For the moment, our framework assumes that those nonlinear relationships are given. An interesting avenue of research would be to develop the mathematics for the learning of those nonlinear relationships, along with the other GMM parameters (such as defined in Appendix \ref{UpdateTheta}).

Our PGI framework is composed of three regularized optimization problems (Fig. \ref{fig:Framework_with_numbers.pdf}). \citet{ggz389} laid the mathematical foundation of the framework, with an emphasis on the inclusion and the learning of the petrophysical signatures (Process $2$ in Fig. \ref{fig:Framework_with_numbers.pdf}). In the current study, we focused on the coupling of several geophysical surveys with various physical properties, thus extending the Process $1$ in Fig. \ref{fig:Framework_with_numbers.pdf}. The third and last process, the geological identification, is still one where there is much room for advancement. \citet{ggz389} showed, with a direct-current resistivity example, the efficacy of the proportions when they are locally set to zero or unity. Proportions values of zero or unity are "constraining," in the sense that they forbid local occurrences of certain units, rather than just favouring it.  While intermediate values of the global proportions appear to have minimal impact on the inversions in our experiments, further studies are required to address the importance and the effects of intermediate (strictly between zero and unity) local proportions. The approach taken by \citet{Giraud2017} combines local proportions from stochastic geological modelling with various warm-started initial models. The integration, extrapolation, and learning of geological information, is part of our current active research. More types of geological information, such as dips, contacts or strikes, also need to be formalized within our framework approach. Combining the PGI smallness term with approaches including prior structural information in the smoothness terms \citep{SparseNorms2, LelievreStructural, BrownStructural, YanStructural, GiraudStructural} is also to be investigated.

This framework has been implemented as part of the open-source \texttt{SimPEG} (Simulation and Parameter Estimation in Geophysics) project \citep{Cockett2015}. This has two major advantages. First, this enables the reproducibility of the approach by making freely available online, in \texttt{GitHub} repositories, the software environment and Python scripts to recreate the example (\url{https://github.com/simpeg-research/Astic-2020-JointInversion}, \citet{PGIJointExamples}). Second, by providing a common environment, it allows the implementation of the framework to be readily used with any type of geophysical surveys (such as EM surveys) or discretization (such as OcTree meshes) that are supported in the source code, and facilitate the collaboration with others researchers \citep{SeogiJointEM}.

\section{Conclusion}

We have expanded the PGI framework to use petrophysical and geological information, represented as a Gaussian Mixture Model, as a coupling term to perform multi-physics joint inversions. We described our strategies for handling multiple geophysical target misfits as well as a petrophysical target misfit. We presented our efforts to make the implementation modular, extensible, and shareable. Finally, we demonstrated, through the DO-$27$ kimberlite pipe synthetic example, the gains that can be made by including various types of information into a single inversion. Only a joint approach for inverting the potential field datasets allowed us to delineate two kimberlite facies and to reproduce their petrophysical signature.

\section{Acknowledgments}

We sincerely thank the open-source software \texttt{SimPEG} community whose work has considerably facilitated the research presented here. Special thanks go to Dominique Fournier, for the implementation of the potential fields operators, Seogi Kang, for contributions to the \texttt{mapping} module \citep{SeogiMapping} and Rowan Cockett for the pioneering development of \texttt{SimPEG}. We thank Condor Consulting Inc., Peregrine Diamonds Ltd. and Kennecott for making the DO-$27$ geology model available for our research.

\bibliographystyle{gji}
\bibliography{bibliography}

\begin{thebibliography}{74}
\expandafter\ifx\csname natexlab\endcsname\relax\def\natexlab#1{#1}\fi

\bibitem[Afnimar et~al.(2002)Afnimar, Koketsu, \& Nakagawa]{Afnimar2002}
Afnimar, P., Koketsu, K., \& Nakagawa, K., 2002.
\newblock Joint inversion of refraction and gravity data for the
  three-dimensional topography of a sediment--basement interface, {\it
  Geophysical Journal International\/}, {\bf 151}(1), 243--254.

\bibitem[Astic(2020)]{PGIJointExamples}
Astic, T., 2020.
\newblock {Collection of scripts for forward modelling and joint inversion of
  potential fields data}, \url{https://doi.org/10.5281/zenodo.3571471},
  Accessed: 2020-01-31.

\bibitem[Astic \& Oldenburg(2019)]{ggz389}
Astic, T. \& Oldenburg, D.~W., 2019.
\newblock {A framework for petrophysically and geologically guided geophysical
  inversion using a dynamic Gaussian mixture model prior}, {\it Geophysical
  Journal International\/}, {\bf 219}(3), 1989--2012.

\bibitem[Bijani et~al.(2017)Bijani, Leli{\`e}vre, Ponte-Neto, \&
  Farquharson]{Bijani2017}
Bijani, R., Leli{\`e}vre, P.~G., Ponte-Neto, C.~F., \& Farquharson, C.~G.,
  2017.
\newblock Physical-property-, lithology- and surface-geometry-based joint
  inversion using pareto multi-objective global optimization, {\it Geophysical
  Journal International\/}, {\bf 209}(2), 730--748.

\bibitem[Bosch(2004)]{Bosch2004}
Bosch, M., 2004.
\newblock {The optimization approach to lithological tomography: Combining
  seismic data and petrophysics for porosity prediction}, {\it Geophysics\/},
  {\bf 69}(5), 1272--1282.

\bibitem[Bournas et~al.(2018)Bournas, Prikhodko, Kwan, Legault, Polianichko, \&
  Treshchev]{Bournas}
Bournas, N., Prikhodko, A., Kwan, K., Legault, J., Polianichko, V., \&
  Treshchev, S., 2018.
\newblock A new approach for kimberlite exploration using helicopter-borne tdem
  data, in {\em SEG Technical Program Expanded Abstracts 2018\/}, pp.
  1853--1857.

\bibitem[Brown et~al.(2012)Brown, Key, \& Singh]{BrownStructural}
Brown, V., Key, K., \& Singh, S., 2012.
\newblock {Seismically regularized controlled-source electromagnetic
  inversion}, {\it Geophysics\/}, {\bf 77}(1), E57--E65.

\bibitem[Chen \& Hoversten(2012)]{Chen2012}
Chen, J. \& Hoversten, G.~M., 2012.
\newblock {Joint inversion of marine seismic AVA and CSEM data using
  statistical rock-physics models and Markov random fields}, {\it
  Geophysics\/}, {\bf 77}(1), R65--R80.

\bibitem[Chen et~al.(2007)Chen, Hoversten, Vasco, Rubin, \& Hou]{Chen2007}
Chen, J., Hoversten, G.~M., Vasco, D., Rubin, Y., \& Hou, Z., 2007.
\newblock {A Bayesian model for gas saturation estimation using marine seismic
  AVA and CSEM data}, {\it Geophysics\/}, {\bf 72}(2), WA85--WA95.

\bibitem[Cockett et~al.(2015)Cockett, Kang, Heagy, Pidlisecky, \&
  Oldenburg]{Cockett2015}
Cockett, R., Kang, S., Heagy, L.~J., Pidlisecky, A., \& Oldenburg, D.~W., 2015.
\newblock {SimPEG: An open source framework for simulation and gradient based
  parameter estimation in geophysical applications}, {\it Computers and
  Geosciences\/}, {\bf 85}, 142--154.

\bibitem[De~Stefano et~al.(2011)De~Stefano, Golfr{\'e}~Andreasi, Re, Virgilio,
  \& Snyder]{Stefano2011}
De~Stefano, M., Golfr{\'e}~Andreasi, F., Re, S., Virgilio, M., \& Snyder,
  F.~F., 2011.
\newblock {Multiple-domain, simultaneous joint inversion of geophysical data
  with application to subsalt imaging}, {\it Geophysics\/}, {\bf 76}(3),
  R69--R80.

\bibitem[Dempster et~al.(1977)Dempster, Laird, \&
  Rubin]{ExpectationMaximization}
Dempster, A.~P., Laird, N.~M., \& Rubin, D.~B., 1977.
\newblock {Maximum likelihood from incomplete data via the EM algorithm}, {\it
  Journal of the Royal Statistical Society, series B\/}, {\bf 39}(1), 1--38.

\bibitem[Devriese et~al.(2017)Devriese, Davis, \& Oldenburg]{TKCpaper}
Devriese, S. G.~R., Davis, K., \& Oldenburg, D.~W., 2017.
\newblock {Inversion of airborne geophysics over the DO-27/DO-18 kimberlites
  --- Part 1: Potential fields}, {\it Interpretation\/}, {\bf 5}(3),
  T299--T311.

\bibitem[Doetsch et~al.(2010)Doetsch, Linde, Coscia, Greenhalgh, \&
  Green]{Doetsch2010}
Doetsch, J., Linde, N., Coscia, I., Greenhalgh, S.~A., \& Green, A.~G., 2010.
\newblock {Zonation for 3D aquifer characterization based on joint inversions
  of multimethod crosshole geophysical data}, {\it Geophysics\/}, {\bf 75}(6),
  G53--G64.

\bibitem[Fournier \& Oldenburg(2019)]{SparseNorms2}
Fournier, D. \& Oldenburg, D.~W., 2019.
\newblock {Inversion using spatially variable mixed {$\ell$}p norms}, {\it
  Geophysical Journal International\/}, {\bf 218}(1), 268--282.

\bibitem[Fournier et~al.(2017)Fournier, Kang, McMillan, \& Oldenburg]{TKCEM}
Fournier, D., Kang, S., McMillan, M.~S., \& Oldenburg, D.~W., 2017.
\newblock {Inversion of airborne geophysics over the DO-27/DO-18 kimberlites
  --- Part 2: Electromagnetics}, {\it Interpretation\/}, {\bf 5}(3),
  T313--T325.

\bibitem[Gallardo \& Meju(2003)]{Gallardo2003}
Gallardo, L.~A. \& Meju, M.~A., 2003.
\newblock {Characterization of heterogeneous near-surface materials by joint 2D
  inversion of dc resistivity and seismic data}, {\it Geophysical Research
  Letters\/}, {\bf 30}(13).

\bibitem[Gallardo \& Meju(2004)]{gallardo_strategy}
Gallardo, L.~A. \& Meju, M.~A., 2004.
\newblock {Joint two-dimensional DC resistivity and seismic travel time
  inversion with cross-gradients constraints}, {\it Journal of Geophysical
  Research: Solid Earth\/}, {\bf 109}(B3).

\bibitem[Gallardo \& Meju(2011)]{Gallardo2011}
Gallardo, L.~A. \& Meju, M.~A., 2011.
\newblock {Structure-coupled multiphysics imaging in geophysical sciences},
  {\it Reviews of Geophysics\/}, {\bf 49}(1).

\bibitem[Giraud et~al.(2017)Giraud, Pakyuz-Charrier, Jessell, Lindsay, Martin,
  \& Ogarko]{Giraud2017}
Giraud, J., Pakyuz-Charrier, E., Jessell, M., Lindsay, M., Martin, R., \&
  Ogarko, V., 2017.
\newblock {Uncertainty reduction in joint inversion using geologically
  conditioned petrophysical constraints}, {\it Geophysics\/}, {\bf 82}(6),
  1--16.

\bibitem[Giraud et~al.(2019{\natexlab{a}})Giraud, Lindsay, Ogarko, Jessell,
  Martin, \& Pakyuz-Charrier]{GiraudStructural}
Giraud, J., Lindsay, M., Ogarko, V., Jessell, M., Martin, R., \&
  Pakyuz-Charrier, E., 2019{\natexlab{a}}.
\newblock {Integration of geoscientific uncertainty into geophysical inversion
  by means of local gradient regularization}, {\it Solid Earth\/}, {\bf 10}(1),
  193--210.

\bibitem[Giraud et~al.(2019{\natexlab{b}})Giraud, Ogarko, Lindsay,
  Pakyuz-Charrier, Jessell, \& Martin]{Giraud2019}
Giraud, J., Ogarko, V., Lindsay, M., Pakyuz-Charrier, E., Jessell, M., \&
  Martin, R., 2019{\natexlab{b}}.
\newblock Sensitivity of constrained joint inversions to geological and
  petrophysical input data uncertainties with posterior geological analysis,
  {\it Geophysical Journal International\/}, {\bf 218}(1), 666--688.

\bibitem[Giuseppe et~al.(2014)Giuseppe, Troiano, Troise, \&
  Natale]{PostInversionClustering0}
Giuseppe, M. G.~D., Troiano, A., Troise, C., \& Natale, G.~D., 2014.
\newblock {k-Means clustering as tool for multivariate geophysical data
  analysis. An application to shallow fault zone imaging}, {\it Journal of
  Applied Geophysics\/}, {\bf 101}, 108 -- 115.

\bibitem[Grana \& {Della Rossa}(2010)]{Grana2010}
Grana, D. \& {Della Rossa}, E., 2010.
\newblock {Probabilistic petrophysical-properties estimation integrating
  statistical rock physics with seismic inversion}, {\it Geophysics\/}, {\bf
  75}(3), O21.

\bibitem[Grana et~al.(2017)Grana, Fjeldstad, \& Omre]{Grana2017}
Grana, D., Fjeldstad, T., \& Omre, H., 2017.
\newblock {Bayesian Gaussian Mixture Linear Inversion for Geophysical Inverse
  Problems}, {\it Mathematical Geosciences\/}, {\bf 49}(4), 493--515.

\bibitem[Haber \& Oldenburg(1997)]{Haber1997}
Haber, E. \& Oldenburg, D., 1997.
\newblock Joint inversion: a structural approach, {\it Inverse Problems\/},
  {\bf 13}(1), 63.

\bibitem[Hansen(2000)]{Hansenlcurve}
Hansen, P.~C., 2000.
\newblock {The L-Curve and its Use in the Numerical Treatment of Inverse
  Problems}, in {\em {in Computational Inverse Problems in Electrocardiology,
  ed. P. Johnston, Advances in Computational Bioengineering}\/}, pp. 119--142,
  WIT Press.

\bibitem[Hansen \& O'Leary(1993)]{HansenLcurve0}
Hansen, P.~C. \& O'Leary, D.~P., 1993.
\newblock {The Use of the L-Curve in the Regularization of Discrete Ill-Posed
  Problems}, {\it SIAM Journal on Scientific Computing\/}, {\bf 14}(6),
  1487--17, Copyright - Copyright] {\copyright} 1993 Society for Industrial and
  Applied Mathematics; Last updated - 2012-06-29.

\bibitem[Heagy et~al.(2017)Heagy, Cockett, Kang, Rosenkjaer, \&
  Oldenburg]{heagy2017framework}
Heagy, L.~J., Cockett, R., Kang, S., Rosenkjaer, G.~K., \& Oldenburg, D.~W.,
  2017.
\newblock A framework for simulation and inversion in electromagnetics, {\it
  Computers \& Geosciences\/}, {\bf 107}, 1--19.

\bibitem[Hoversten et~al.(2006)Hoversten, Cassassuce, Gasperikova, Newman,
  Chen, Rubin, Hou, \& Vasco]{Hoversten2006}
Hoversten, G.~M., Cassassuce, F., Gasperikova, E., Newman, G.~A., Chen, J.,
  Rubin, Y., Hou, Z., \& Vasco, D., 2006.
\newblock {Direct reservoir parameter estimation using joint inversion of
  marine seismic AVA and CSEM data}, {\it Geophysics\/}, {\bf 71}(3), C1--C13.

\bibitem[Jansen \& Witherly(2004)]{JansenEtAl2004}
Jansen, J. \& Witherly, K., 2004.
\newblock The {T}li {K}wi {C}ho kimberlite complex, {N}orthwest {T}erritories,
  {C}anada: {A} geophysical case study, in {\em 2004 SEG Annual Meeting\/}.

\bibitem[Jegen et~al.(2009)Jegen, Hobbs, Tarits, \& Chave]{Jegen2009}
Jegen, M.~D., Hobbs, R.~W., Tarits, P., \& Chave, A., 2009.
\newblock {Joint inversion of marine magnetotelluric and gravity data
  incorporating seismic constraints: Preliminary results of sub-basalt imaging
  off the Faroe Shelf}, {\it Earth and Planetary Science Letters\/}, {\bf
  282}(1-4), 47--55.

\bibitem[Kamm et~al.(2015)Kamm, Lundin, Bastani, Sadeghi, \&
  Pedersen]{Kamm2015}
Kamm, J., Lundin, I.~A., Bastani, M., Sadeghi, M., \& Pedersen, L.~B., 2015.
\newblock {Joint inversion of gravity, magnetic, and petrophysical data --- A
  case study from a gabbro intrusion in Boden, Sweden}, {\it Geophysics\/},
  {\bf 80}(5), B131--B152.

\bibitem[Kang et~al.(2015)Kang, Cockett, Heagy, \& Oldenburg]{SeogiMapping}
Kang, S., Cockett, R., Heagy, L.~J., \& Oldenburg, D.~W., 2015.
\newblock {Moving between dimensions in electromagnetic inversions}, in {\em
  {SEG Technical Program Expanded Abstracts 2015}\/}, pp. 5000--5004.

\bibitem[Kang et~al.(2017)Kang, Fournier, \& Oldenburg]{TKCIP}
Kang, S., Fournier, D., \& Oldenburg, D.~W., 2017.
\newblock {Inversion of airborne geophysics over the DO-27/DO-18 kimberlites
  --- Part 3: Induced polarization}, {\it Interpretation\/}, {\bf 5}(3),
  T327--T340.

\bibitem[Keating \& Sailhac(2004)]{Keating}
Keating, P. \& Sailhac, P., 2004.
\newblock {Use of the analytic signal to identify magnetic anomalies due to
  kimberlite pipes}, {\it Geophysics\/}, {\bf 69}(1), 180--190.

\bibitem[Leli{\`e}vre \& Farquharson(2016)]{Lelievre2016}
Leli{\`e}vre, P.~G. \& Farquharson, C.~G., 2016.
\newblock {Integrated Imaging for Mineral Exploration}, in {\em {Integrated
  Imaging of the Earth}\/}, chap.~8, pp. 137--166, American Geophysical Union
  (AGU).

\bibitem[Leli{\`e}vre \& Oldenburg(2009)]{LelievreStructural}
Leli{\`e}vre, P.~G. \& Oldenburg, D.~W., 2009.
\newblock {A comprehensive study of including structural orientation
  information in geophysical inversions}, {\it Geophysical Journal
  International\/}, {\bf 178}(2), 623--637.

\bibitem[Leli{\`e}vre et~al.(2009)Leli{\`e}vre, Oldenburg, \&
  Williams]{Lelievre2009}
Leli{\`e}vre, P.~G., Oldenburg, D.~W., \& Williams, N.~C., 2009.
\newblock {Integrating geological and geophysical data through advanced
  constrained inversions}, {\it Exploration Geophysics\/}, {\bf 40}(4),
  334--341.

\bibitem[Leli{\`e}vre et~al.(2012)Leli{\`e}vre, Farquharson, \&
  Hurich]{Lelievre2012}
Leli{\`e}vre, P.~G., Farquharson, C.~G., \& Hurich, C.~A., 2012.
\newblock {Joint inversion of seismic traveltimes and gravity data on
  unstructured grids with application to mineral exploration}, {\it
  Geophysics\/}, {\bf 77}(1), K1--K15.

\bibitem[Li \& Oldenburg(1996)]{Li1996}
Li, Y. \& Oldenburg, D.~W., 1996.
\newblock {3-D inversion of magnetic data}, {\it Geophysics\/}, {\bf 61}(2),
  394--408.

\bibitem[Li \& Oldenburg(1998)]{Li1998}
Li, Y. \& Oldenburg, D.~W., 1998.
\newblock {3-D inversion of gravity data}, {\it Geophysics\/}, {\bf 63}(1),
  109--119.

\bibitem[Li et~al.(2019)Li, Melo, Martinez, \& Sun]{QuasiGeologicalModel}
Li, Y., Melo, A., Martinez, C., \& Sun, J., 2019.
\newblock {Geology differentiation: A new frontier in quantitative geophysical
  interpretation in mineral exploration}, {\it The Leading Edge\/}, {\bf
  38}(1), 60--66.

\bibitem[Macnae(1995)]{Macnae}
Macnae, J., 1995.
\newblock {Applications of geophysics for the detection and exploration of
  kimberlites and lamproites}, {\it Journal of Geochemical Exploration\/}, {\bf
  53}(1), 213 -- 243, Diamond Exploration: Into the 21st Century.

\bibitem[Martinez \& Li(2015)]{PostInversionClustering2}
Martinez, C. \& Li, Y., 2015.
\newblock {Lithologic characterization using airborne gravity gradient and
  aeromagnetic data for mineral exploration: A case study in the
  Quadril{\'a}tero Ferr{\'\i}fero, Brazil}, {\it Interpretation\/}, {\bf 3}(2),
  SL1--SL13.

\bibitem[Mehanee et~al.(2005)Mehanee, Golubev, \& Zhdanov]{SensW}
Mehanee, S., Golubev, N., \& Zhdanov, M.~S., 2005.
\newblock {Weighted regularized inversion of magnetotelluric data}, in {\em
  {SEG Technical Program Expanded Abstracts 1998}\/}, pp. 481--484.

\bibitem[Meju \& Gallardo(2016)]{Meju2016}
Meju, M.~A. \& Gallardo, L.~A., 2016.
\newblock {Structural Coupling Approaches in Integrated Geophysical Imaging},
  in {\em {Integrated Imaging of the Earth}\/}, chap.~4, pp. 49--67, American
  Geophysical Union (AGU).

\bibitem[Melo et~al.(2017)Melo, Sun, \& Li]{Melo2017}
Melo, A.~T., Sun, J., \& Li, Y., 2017.
\newblock {Geophysical inversions applied to 3D geology characterization of an
  iron oxide copper-gold deposit in Brazil}, {\it Geophysics\/}, {\bf 82}(5),
  K1--K13.

\bibitem[Moorkamp et~al.(2011)Moorkamp, Heincke, Jegen, Roberts, \&
  Hobbs]{Moorkamp2011}
Moorkamp, M., Heincke, B., Jegen, M., Roberts, A.~W., \& Hobbs, R.~W., 2011.
\newblock {A framework for 3-D joint inversion of MT, gravity and seismic
  refraction data}, {\it Geophysical Journal International\/}, {\bf 184}(1),
  477--493.

\bibitem[Moorkamp et~al.(2016{\natexlab{a}})Moorkamp, Heincke, Jegen, Hobbs, \&
  Roberts]{doi:10.1002/9781118929063.ch9}
Moorkamp, M., Heincke, B., Jegen, M., Hobbs, R.~W., \& Roberts, A.~W.,
  2016{\natexlab{a}}.
\newblock Joint inversion in hydrocarbon exploration, in {\em Integrated
  Imaging of the Earth\/}, chap.~9, pp. 167--189, American Geophysical Union
  (AGU).

\bibitem[Moorkamp et~al.(2016{\natexlab{b}})Moorkamp, Leli{\`e}vre, Linde, \&
  Khan]{integratedImaging}
Moorkamp, M., Leli{\`e}vre, P.~G., Linde, N., \& Khan, A., 2016{\natexlab{b}}.
\newblock {\it {Integrated Imaging of the Earth: Theory and Applications}\/},
  American Geophysical Union (AGU).

\bibitem[Murphy(2012)]{Murphy2012}
Murphy, K.~P., 2012.
\newblock {\it Machine Learning: A Probabilistic Perspective\/}, The MIT Press.

\bibitem[Oldenburg \& Li(2005)]{Tutorial}
Oldenburg, D.~W. \& Li, Y., 2005.
\newblock {Inversion for Applied Geophysics: A Tutorial}, in {\em {Near-Surface
  Geophysics}\/}, chap.~5, pp. 89--150, ed. Butler, D.~K., Society of
  Exploration Geophysicists.

\bibitem[Oldenburg et~al.(1997)Oldenburg, Li, \& Ellis]{OldenburgMilligan1997}
Oldenburg, D.~W., Li, Y., \& Ellis, R.~G., 1997.
\newblock {Inversion of geophysical data over a copper gold porphyry deposit: A
  case history for Mt. Milligan}, {\it Geophysics\/}, {\bf 62}(5), 1419--1431.

\bibitem[Oldenburg et~al.(2019)Oldenburg, Heagy, Kang, \&
  Cockett]{SeogiJointEM}
Oldenburg, D.~W., Heagy, L.~J., Kang, S., \& Cockett, R., 2019.
\newblock {3D electromagnetic modelling and inversion: a case for open source},
  {\it Exploration Geophysics\/}, {\bf 0}(0), 1--13.

\bibitem[Onizawa et~al.(2002)Onizawa, Mikada, Watanabe, \&
  Sakashita]{Onizawa2002}
Onizawa, S., Mikada, H., Watanabe, H., \& Sakashita, S., 2002.
\newblock {A method for simultaneous velocity and density inversion and its
  application to exploration of subsurface structure beneath Izu-Oshima
  volcano, Japan}, {\it Earth, Planets and Space\/}, {\bf 54}(8), 803--817.

\bibitem[Paasche(2016)]{PostInversionClustering3}
Paasche, H., 2016.
\newblock {Post-Inversion Integration of Disparate Tomographic Models by Model
  Structure Analyses}, in {\em {Integrated Imaging of the Earth}\/}, chap.~5,
  pp. 69--91, American Geophysical Union (AGU).

\bibitem[Paasche \& Tronicke(2007)]{Paasche2007}
Paasche, H. \& Tronicke, J., 2007.
\newblock {Cooperative inversion of 2D geophysical data sets: A zonal approach
  based on fuzzy c-means cluster analysis}, {\it Geophysics\/}, {\bf 72}(3),
  A35--A39.

\bibitem[Paasche et~al.(2006)Paasche, Tronicke, Holliger, Green, \&
  Maurer]{PostInversionClustering1}
Paasche, H., Tronicke, J., Holliger, K., Green, A.~G., \& Maurer, H., 2006.
\newblock {Integration of diverse physical-property models: Subsurface zonation
  and petrophysical parameter estimation based on fuzzy c -means cluster
  analyses}, {\it Geophysics\/}, {\bf 71}(3), H33--H44.

\bibitem[Parker(1977)]{Parker}
Parker, R.~L., 1977.
\newblock {Understanding Inverse Theory}, {\it Annual Review of Earth and
  Planetary Sciences\/}, {\bf 5}(1), 35--64.

\bibitem[Pearson(1900)]{Pearson1900}
Pearson, K., 1900.
\newblock {On the criterion that a given system of deviations from the probable
  in the case of a correlated system of variables is such that it can be
  reasonably supposed to have arisen from random sampling}, {\it The London,
  Edinburgh, and Dublin Philosophical Magazine and Journal of Science\/}, {\bf
  50}(302), 157--175.

\bibitem[Pedregosa et~al.(2011)Pedregosa, Varoquaux, Gramfort, Michel, Thirion,
  Grisel, Blondel, Prettenhofer, Weiss, Dubourg, Vanderplas, Passos,
  Cournapeau, Brucher, Perrot, \& Duchesnay]{scikit-learn}
Pedregosa, F., Varoquaux, G., Gramfort, A., Michel, V., Thirion, B., Grisel,
  O., Blondel, M., Prettenhofer, P., Weiss, R., Dubourg, V., Vanderplas, J.,
  Passos, A., Cournapeau, D., Brucher, M., Perrot, M., \& Duchesnay, E., 2011.
\newblock {Scikit-learn: Machine Learning in Python}, {\it Journal of Machine
  Learning Research\/}, {\bf 12}, 2825--2830.

\bibitem[Portniaguine \& Zhdanov(2002)]{Portniaguine2002}
Portniaguine, O. \& Zhdanov, M.~S., 2002.
\newblock {3‐D magnetic inversion with data compression and image focusing},
  {\it Geophysics\/}, {\bf 67}(5), 1532--1541.

\bibitem[Santos \& Bassrei(2007)]{SantosLcurve}
Santos, E. \& Bassrei, A., 2007.
\newblock {L- and {$\Theta$}-curve approaches for the selection of
  regularization parameter in geophysical diffraction tomography}, {\it
  Computers \& Geosciences\/}, {\bf 33}(5), 618 -- 629.

\bibitem[Shamsipour et~al.(2012)Shamsipour, Marcotte, \&
  Chouteau]{Shamsipour2012}
Shamsipour, P., Marcotte, D., \& Chouteau, M., 2012.
\newblock {3D stochastic joint inversion of gravity and magnetic data}, {\it
  Journal of Applied Geophysics\/}, {\bf 79}, 27 -- 37.

\bibitem[Sosa et~al.(2013)Sosa, Velasco, Velazquez, Argaez, \& Romero]{Sosa}
Sosa, A., Velasco, A.~A., Velazquez, L., Argaez, M., \& Romero, R., 2013.
\newblock {Constrained optimization framework for joint inversion of
  geophysical data sets}, {\it Geophysical Journal International\/}, {\bf
  195}(3), 1745--1762.

\bibitem[Sun \& Li(2015)]{Sun2015}
Sun, J. \& Li, Y., 2015.
\newblock {Multidomain petrophysically constrained inversion and geology
  differentiation using guided fuzzy c-means clustering}, {\it Geophysics\/},
  {\bf 80}(4), ID1.

\bibitem[Sun \& Li(2016)]{Sun2016}
Sun, J. \& Li, Y., 2016.
\newblock {Joint inversion of multiple geophysical data using guided fuzzy
  c-means clustering}, {\it Geophysics\/}, {\bf 81}(3), ID37--ID57.

\bibitem[Sun \& Li(2017)]{Sun2017}
Sun, J. \& Li, Y., 2017.
\newblock {Joint inversion of multiple geophysical and petrophysical data using
  generalized fuzzy clustering algorithms}, {\it Geophysical Journal
  International\/}, {\bf 208}(2), 1201--1216.

\bibitem[Tarantola(2005)]{Tarantola}
Tarantola, A., 2005.
\newblock {\it {Inverse Problem Theory and Methods for Model Parameter
  Estimation}\/}, vol.~89, siam.

\bibitem[Tikhonov \& Arsenin(1977)]{tikhonov1977solutions}
Tikhonov, A.~N. \& Arsenin, V.~Y., 1977.
\newblock {\it Solutions of ill-posed problems\/}, V. H. Winston \& Sons,
  Washington, D.C.: John Wiley \& Sons, New York, Translated from the Russian,
  Preface by translation editor Fritz John, Scripta Series in Mathematics.

\bibitem[Williams(2006)]{williams2006applying}
Williams, N.~C., 2006.
\newblock {Applying UBC-GIF potential field inversions in greenfields or
  brownfields exploration}, in {\em Proceedings of the Australian Earth
  Sciences Convention. ASEG/GSA.\/}.

\bibitem[Williams(2008)]{Williams_2008}
Williams, N.~C., 2008.
\newblock {\it Geologically-constrained UBC-GIF gravity and magnetic inversions
  with examples from the Agnew-Wiluna greenstone belt, Western Australia\/},
  Ph.D. thesis, University of British Columbia.

\bibitem[Yan et~al.(2017)Yan, Kalscheuer, Hedin, \&
  Garcia~Juanatey]{YanStructural}
Yan, P., Kalscheuer, T., Hedin, P., \& Garcia~Juanatey, M.~A., 2017.
\newblock {Two-dimensional magnetotelluric inversion using reflection seismic
  data as constraints and application in the COSC project}, {\it Geophysical
  Research Letters\/}, {\bf 44}(8), 3554--3563.

\end{thebibliography}

\appendix

\section{Working with nonlinear petrophysical relationships} \label{LinearWithMappingSection}

\begin{figure}
    \includegraphics[width=\columnwidth]{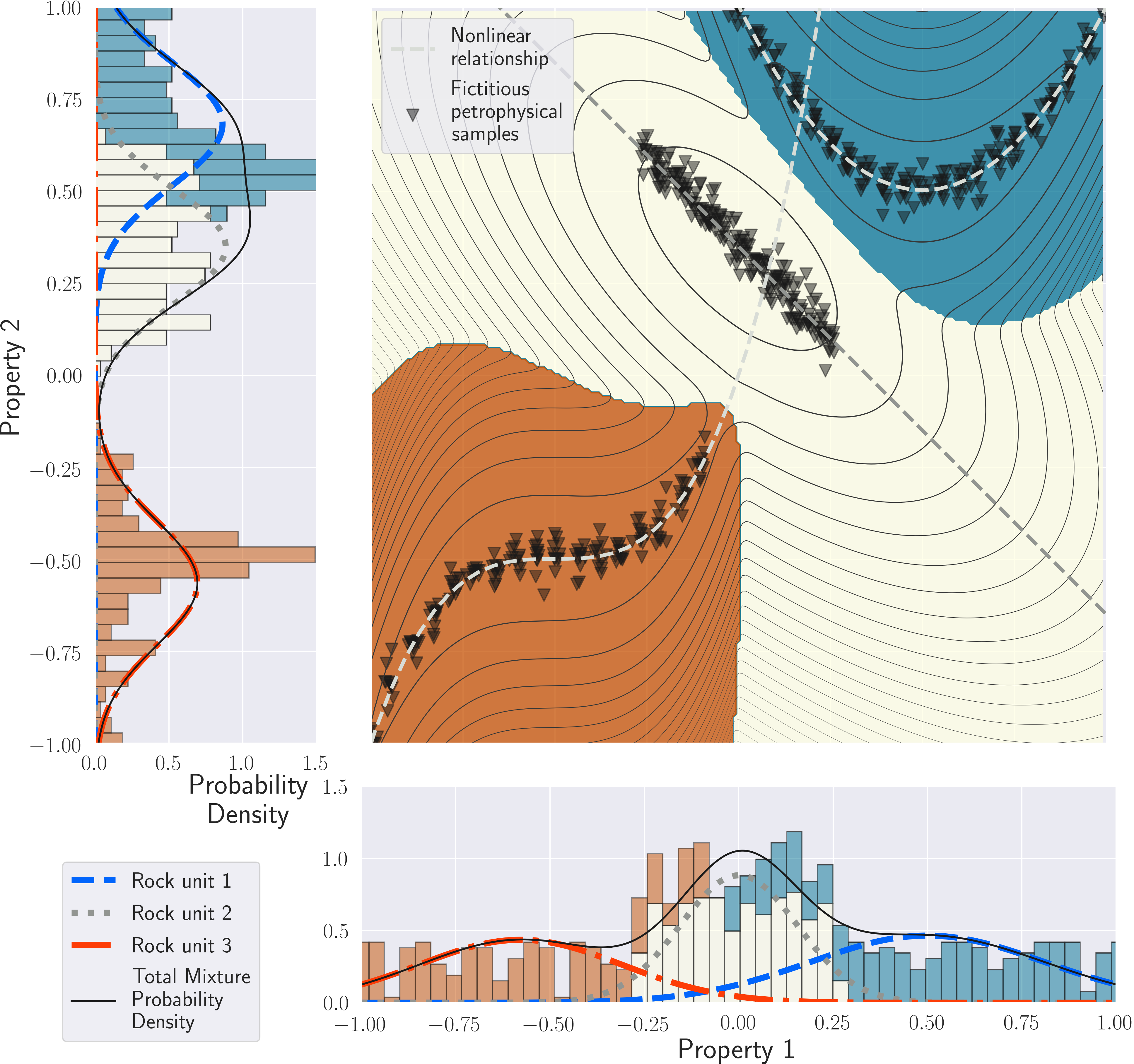}
    \caption{GMM with various polynomial relationships: one linear (no addition required), one quadratic and one cubic. In the main panel, thicker contour lines indicate higher probabilities. The 1D probability distribution for each physical property, and the respective histogram of each unit, are projected on the left and bottom panels.}
    \label{fig:example_projection.png}
\end{figure}

While linear trends can be accounted through the covariance matrix to obtain elongated clusters, it is also possible to account for nonlinear relationships between physical properties in our framework (Fig. \ref{fig:example_projection.png}). This is achieved by composing the Gaussian function with the nonlinear relationship $P_j$ between the physical properties of the particular rock unit $j$. This corresponds to using $\mathcal{N}(P_j(\mathbf{m}_i)|\mitbf{\mu}_j, \mathbf{W}_i^{-1}\mitbf{\Sigma}_j\mathbf{W}_i^{-1})$ for each rock unit $j$ in the GMM in equation \eqref{eq:mixturemodel}.

In this section, we present a joint inversion of two linear problems whose respective physical properties have nonlinear petrophysical relationships between each other. For simplicity, the physics of the two problems is the same and is based on the example previously used in \citet{Li1996}. The models are discretized on a 1D mesh defined on the interval $[0, 1]$ and divided into 100 cells. Both datasets consist of $30$ data points at various frequencies equally distributed from $1$ to $59$ evaluated according to:
\begin{equation}
d_j = \int_0^1 e^{-jx}\cos(2\pi j x) m(x) dx, ~j=1, 3, .., 59.
\end{equation}

Each model presents three distinct units. The background for both models is set to $0$ while the two other units are linked through a quadratic and a cubic relationship respectively.

\begin{figure*}
    \includegraphics[width=\textwidth]{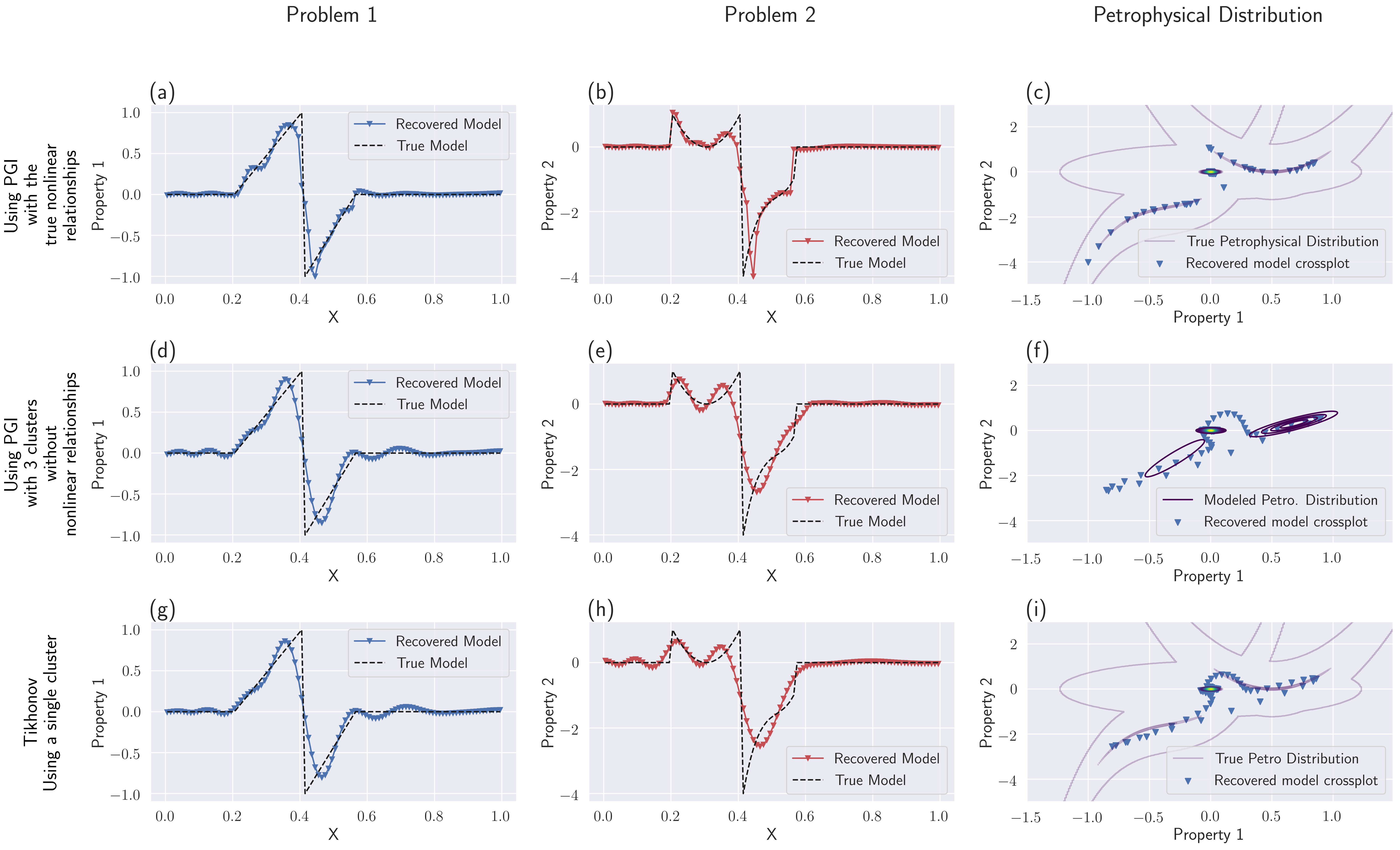}
    \caption{Linear joint inversions with various types of physical properties relationships (no trend, quadratic, cubic). Panels (a) to (c) show the inversion result with PGI using the known nonlinear relationships. The first panel (a) shows the result for the first problem, (b) for the second problem. The panel (c) shows the cross-plot of the models over the contour of the GMM with nonlinear relationships. Panels (d) to (f) show the result with PGI without nonlinear relationships. In panel (f), we show the used GMM without nonlinear relationships. Panels (g) to (i) show the Tikhonov inversions result.}
    \label{fig:LinearWithMapping.png}
\end{figure*}

We invert the two datasets using the multi-physics PGI framework, including \textit{a priori} knowledge of the petrophysical distributions and relationships. The result can be seen in the first row of Figs \ref{fig:LinearWithMapping.png}(a) to (c). Note that the petrophysical relationships are well reproduced. This allows the recovery of details of the two models.

For comparison, we also jointly invert the datasets but without the polynomial relationships (by merely fitting a Gaussian distribution to each unit). The result can be seen in the second row of Figs \ref{fig:LinearWithMapping.png}(d) to (f). While the overall structures are well recovered, we miss some details of the models. The background is not as flat as with the full information, the lower tip of the model of problem $2$ is completely missed.

We also invert both datasets individually using the classic Tikhonov inversion. The result is shown in the last row of Figs \ref{fig:LinearWithMapping.png}(g) to (i). Results are smoother, as expected. The background presents even more variations, but the overall structures are recovered. Same as for the joint inversion without the polynomial relationships, the details are missing.

\section{Updating the Gaussian mixture model in multi-dimensions} \label{UpdateTheta}

In this section, we generalize the learning of the GMM parameters presented in \citet{ggz389} to multivariate Gaussian distributions, representing multiple physical properties. The main difference comes from the fact that the means are now vectors (they are scalars in 1-D) and that the scalar variances in 1D become covariance matrices. We define the problem as a Maximum A Posteriori (MAP) estimate of the GMM means, proportions and covariance matrices, using the MAP Expectation-Maximization (MAP-EM) algorithm \citep{ExpectationMaximization}.
\begin{equation}
\mathcal{P}(\Theta|\mathbf{m}) \propto \mathcal{M}(\mathbf{m}|\Theta)\mathcal{P}(\Theta).
\label{theta_posterior}
\end{equation}

We choose to follow a semi-conjugate prior approach for the choice of the prior distributions $\mathcal{P}(\Theta)$ \citep{ggz389, Murphy2012}. 

The computation of the responsibilities for the E-step of the MAP-EM algorithm stay the same as in \citet{ggz389}, except for the dimension of the parameters:

\begin{equation}
n_{ij}^{(k)} = \frac{\mathcal{P}(z_i=j)^{(k-1)}\mathcal{N}(\mathbf{m}_i|{\mitbf{\mu}_j}^{(k-1)}, {\mitbf{\Sigma}_j}^{(k-1)})}{ \sum_{t=1}^c \mathcal{P}(z_i=t)^{(k-1)} \mathcal{N}({\mathbf{m}}_i|{\mitbf{\mu}_t}^{(k-1)}, {\mitbf{\Sigma}_j}^{(k-1)})} \label{eq:responsibilities}.
\end{equation}

The update to the proportions stays similar to the univariate case \citep{ggz389}:
\begin{align}
&{\pi}^{(k)}_j = \frac{V_{j}^{(k)}+\zeta_j {{\pi}_j}_{\text{prior}}V}{V(1+\sum_{t=1}^c \zeta_t {{\pi}_t}_{\text{prior}})} \label{eq:pi_update}, \\
&\text{with:} \nonumber\\
&V_{j}^{(k)} = \sum^n_{i=1} v_i n_{ij}^{(k)}, \label{VolumeProportions}\\
&\text{ and } V=\sum^n_{i=1} v_i,
\end{align}
where $\zeta_j$ is the confidence in the prior proportion ${\pi_j}_{\text{prior}}$ of the rock unit $j$, $v_i$ is the volume of the $i$\textsuperscript{th} cell and $V$ is the volume of the
active mesh. This allows the estimates to be mesh-independent by using volumetric proportions instead of cell counts.


The mean update proposed in \citet{ggz389} generalizes to each physical property $p$:
\begin{align}
&{\mu_j^p}^{(k)}=\frac{V_{j}^{(k)}{\bar{{m}^p_j}}^{(k)} + \kappa^p_j {\pi_j}_{\text{prior}} V {\mu^p_j}_{\text{prior}}}{V_{j}^{(k)}+\kappa_j^p {\pi_j}_{\text{prior}} V} \label{eq:mu_update},\\
&\text{with:} \nonumber\\
&{\bar{\mathbf{m}}}_j^{(k)} = \frac{\sum^n_{i=1} v_i n_{ij}^{(k)} \mathbf{m}_i}{V_{j}^{(k)}},
\end{align}
where $\kappa_j^p$ is the confidence in ${{\mu^p_j}_{\text{prior}}}$, which is the prior mean of the physical property $p$ of the rock unit $j$. The confidences $\{\mitbf{\kappa}\}$ are thus consider as vectors. This is an important tool that we use in section \ref{sec:InterpreterAssumption} to formulate a geologic assumption about the model.

The update to the covariance matrices is, with $\nu_j$ the confidence in the prior covariance matrix ${\mitbf{\Sigma}_j}_{\text{prior}}$ of the rock unit $j$:
\begin{align}
&{\mitbf{\Sigma}_j}^{(k)} = \frac{{{V_{j}^{(k)}} {\mitbf{\Sigma}_{\bar{\mathbf{m}}}}_j}^{(k)} + \nu_j {\pi_j}_{prior} V {\mitbf{\Sigma}_j}_{prior}}
{{V_{j}^{(k)}} + \nu_j {\pi_j}_{prior} V} \label{eq:sig_update},\\
&\text{with:} \nonumber\\
&{\mitbf{\Sigma}_{\bar{\mathbf{m}}}}_j^{(k)} =\frac{1}{{V_{j}^{(k)}}} \sum_{i=1}^{n} v_i n_{ij}^{(k)}(\mathbf{m}_i-\bar{\mathbf{m}_j}^{(k)})(\mathbf{m}_i-\bar{\mathbf{m}_j}^{(k)})^\top,
\end{align}

\section{Algorithm} \label{algorithm}

\begin{algorithm*}
    \small
    \SetKwInOut{Initialization}{Initialization}
    \caption{PGI extended from \citet{ggz389} for joint inversion}
    \label{algo:algorithm}
    \nl \Initialization{
    \begin{itemize}
    \item \underline{Input:}
    \begin{itemize}
        \item Initial geophysical model $\mathbf{m}^{(0)}$, GMM parameters $\Theta^{(0)}$ and geological model $\mathbf{z}^{(0)}$.
    \end{itemize}

    \item \underline{Parameters:}
    \begin{itemize}
      \item \textit{Objective function}: data's noise $\left\{{\mathbf{W}_d}_p\right\}_{p=1..q}$, $\beta^{(0)}$ and $\left\{\alpha\right\}$.parameters.
      \item \textit{Localized prior}: specific $\mathcal{P}(z_i)$ for available locations $i \in \{1..n\}$, weights $\left\{\mathbf{w}\right\}$.
      \item \textit{GMM prior weights}: $\left\{\mitbf{\kappa}_j, \nu_j, \zeta_j\right\}_{j=1..c}$ for the means, variances and proportions.
      \item \textit{Optimization}: $\beta$-cooling factor $\gamma$ ($> 1$), sufficient decrease rate $\tau$ ($\leq 1$), tolerance on target misfit $\epsilon$.
    \end{itemize}

    \item \underline{Output:}
    \begin{itemize}
        \item $\mathbf{m}$, $\Theta$, $\mathbf{z}$.
    \end{itemize}

    \end{itemize}
    }

    \While{$\text{any}({{\Phi_d^k}}>{{\Phi_d^k}^*}, k=1..r)$ and $\Phi_{\text{petro}}>\Phi_{\text{petro}}^*$}{

        \underline{Objective Function Descent Step:}
        \begin{itemize}
        \item Compute a model perturbation $\delta \mathbf{m}$ using an inexact Gauss-Newton.
        \item Line search with Wolfe condition to find an $\eta$ that satisfy a sufficient decrease of $\Phi$.
        \item Return $\mathbf{m}^{(t)} = \mathbf{m}^{(t-1)}+\eta \delta \mathbf{m}$.
        \end{itemize}

        \nl \underline{Update Petrophysical Distribution}
        \begin{itemize}
        \item Fit a new GMM $\Theta^{(t)}$ on $\mathbf{m}^{(t)}$ such as in equations \eqref{eq:pi_update}, \eqref{eq:mu_update} and \eqref{eq:sig_update} until no sufficient increase of the posterior is observed.
        \end{itemize}

        \nl \underline{Classification:}
        \begin{itemize}
        \item Compute the membership $\mathbf{z}^{(t)}$ of the current model $\mathbf{m}^{(t)}$ as in equation \eqref{eq:membership} using $\Theta^{(t)}$.
        \item Update $\mathbf{m}_{\text{ref}}$ and $\mathbf{W}_s$ according to equations \eqref{eq:mref_update} and \eqref{eq:Ws_update} respectively using $\mathbf{z}^{(t)}$.
        \end{itemize}

        \nl \underline{Update regularization weights:}\\
        \uIf{$\text{all}({\Phi_d^k}^{(t)}\geq\text{max}((1+\epsilon){\Phi_d^k}^*, \tau {\Phi_d^k}^{(t-1)}), k=1..r)$}{
        Decrease $\beta$: $\beta^{(t)}=\frac{\beta^{(t-1)}}{\gamma}$.
        }
        \uElseIf{$\text{all}({\Phi_d^k}^{(t)}\leq{\Phi_d^k}^*, k=1..r)$ and $\Phi_{\text{petro}}>\Phi_{\text{petro}}^*$}{
              Increase $\alpha_s$: $\alpha_s^{(t)}=\alpha_s^{(t-1)} \times \text{median}(\frac{{{\Phi_d^k}^*}}{{{{{\Phi_d^k}}}^{(t)}}}, k=1..r)$ (equation \eqref{eq:alpha_warm}).
        }
        \uIf{(optional) $\text{all}({\Phi_d^k}^{(t)}\leq{\Phi_d^k}^*)$ and $\Phi_{\text{petro}}>\Phi_{\text{petro}}^*$ and $\mathbf{z}^{(t)}==\mathbf{z}^{(t-1)}$}{
        Include $\mathbf{m}_{\text{ref}}$ in Smoothness.
        }

        \nl \underline{Update geophysical data misfit weights:}\\
        \uIf{$\text{any}({\Phi_d^k}^{(t)}\leq {\Phi_d^k}^*, k=1..r)$}{
          update $\left\{\chi\right\}$ according to equations \eqref{eq:chi_update} to \eqref{eq:chi_normalizing}.
        }

    }
\end{algorithm*}

\section{Pseudo-code for the implementation in SimPEG} \label{pseudocode}

Here, we provide an overview for the use of our implementation of the multi-physics PGI framework by other scientists. We outline the main components of the multi-physics PGI implementation using \texttt{SimPEG} and other core tools in the Python ecosystem. As in the paper, we consider the multi-physics inversion of gravity and magnetic data.

We begin by creating a simulation \texttt{mesh} and defining mappings that translate the model vector, which contains all of the parameters we will invert for, to physical properties on the mesh to be used in each forward simulation. The inversion model is a single vector; for an inversion with multiple physical parameters, we stack them and use the \texttt{Wires} map to keep track of which indices in the vector correspond to each physical parameter.
\usestyle{default}
\includecode{model_mappings.py}



Note that mappings can be composed; for example if we wanted to invert for log-susceptibility rather than susceptibility, then we would compose the \texttt{wires.susceptibility} mapping with an \texttt{ExpMap} instance. For examples, see \citet{SeogiMapping}.

Next, we construct the forward simulations for both the gravity and magnetic data. The \texttt{survey} objects contain the locations of the receivers as well as parameters defining the source field for the magnetics simulation (magnitude, inclination, and declination).
\usestyle{default}
\includecode{data_misfits.py}







With both gravity and magnetic simulations defined, we now have the ability to compute predicted data given a model. The next step is to construct the data misfit term as described in equation \eqref{eq:datamisfit}.
\usestyle{default}
\includecode{combodatamisfit.py}



The \texttt{mag\_data} and \texttt{grav\_data} objects contain the observed data as well as user-specified uncertainties, and the scalar \texttt{chi}-values are initialized according to equation \eqref{eq:chi_normalizing}.

Next, we construct the GMM and regularization which consists of the petrophysical smallness term described in equation \eqref{eq:smallness_petro} and smoothness terms for both the density and susceptibility. To construct the GMM, we use the \texttt{WeightedGaussianMixtureModel} object, which inherits from and extends the \texttt{sklearn.mixture.GaussianMixture} object in Scikit-Learn \citep{scikit-learn}. The instantiated \texttt{gmm} object has methods for fitting a GMM on a model and computing membership of a model as needed in steps 4 and 5 in Algorithm \ref{algo:algorithm}.
\usestyle{default}
\includecode{regularization.py}








Having defined the components of the objective function, we now specify the optimization routine, in this case, inexact Gauss Newton with projections for bound constraints.
\usestyle{default}
\includecode{optimization.py}




Finally, we assemble the inversion using \texttt{directives} in SimPEG to orchestrate updates throughout the inversion. Each directive has an \texttt{initialize} method which is called at the beginning of the inversion and can be used to set initial values of parameters, for example to initialize $\beta$ and the $\alpha$-values, as well as an \texttt{end\_iteration} method which is called after a model update (at the end of step 3 in Algorithm \ref{algo:algorithm}) and can be used to update parameters in the inversion. The updates in steps 4 through 7 in Algorithm \ref{algo:algorithm} make use of the \texttt{directives} functionality.
\usestyle{default}
\includecode{inverse_problem.py}




\section{Additional models obtained by PGIs for the DO-$27$ synthetic case study} \label{sec:additionalIndividual}

\begin{figure}
\centering
\includegraphics[width=0.5\textwidth]{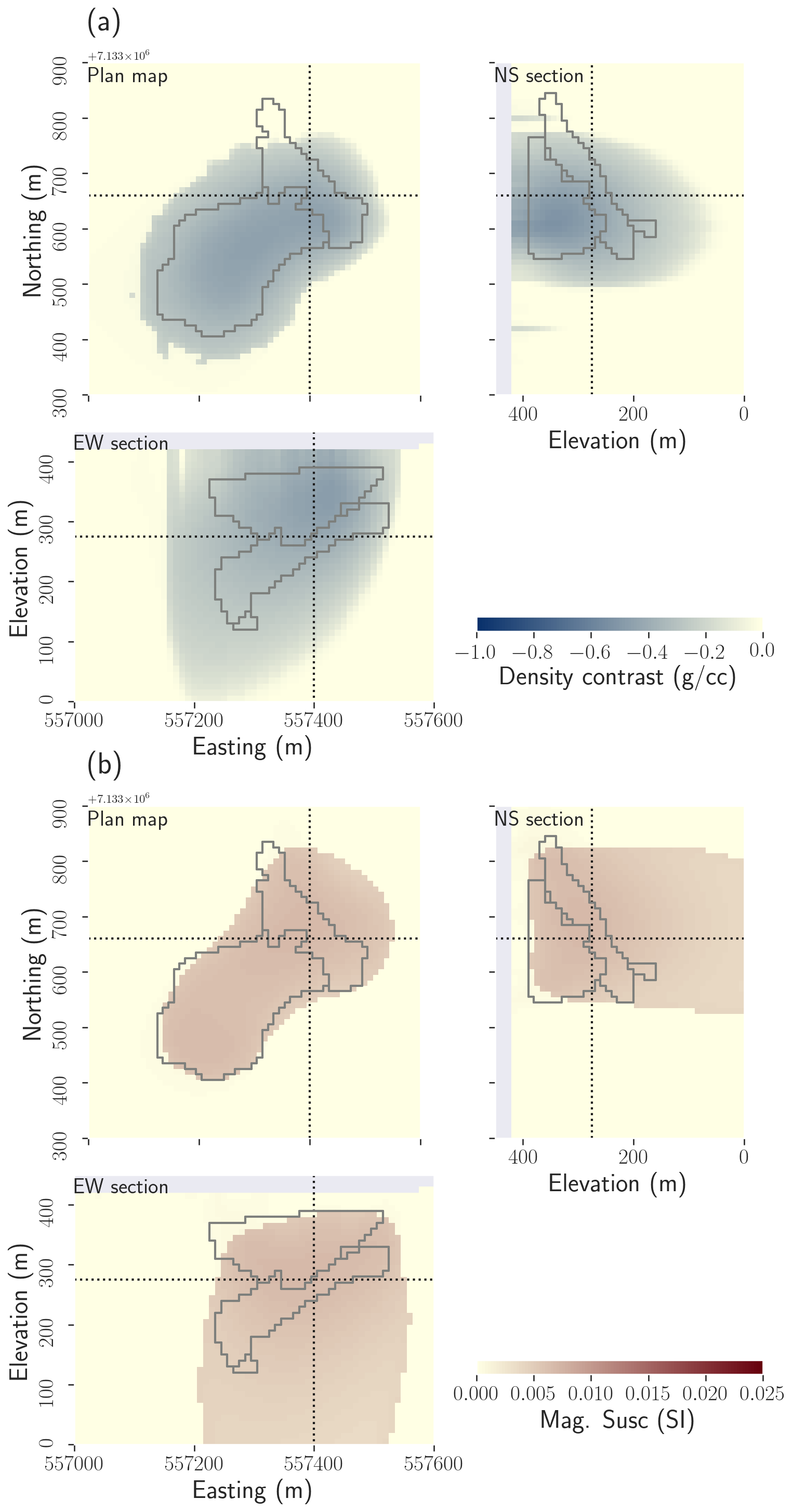}
\caption{(a) magnetic PGI using PK/VK magnetic signature. Note that the recovered volume is much larger than the volume of PK/VK recovered from the gravity inversion with this unit density contrast (Fig. \ref{fig:TKC_IndividualPetro_Synthetic.png}c); (b) gravity PGI using the HK density contrast signature. Note again the larger volume compared to Fig. \ref{fig:TKC_IndividualPetro_Synthetic.png}(d).}
\label{fig:TKC_AdditionalIndividualPetro_FULL_Synthetic.png}
\end{figure}


\begin{figure*}
    \centering
    \includegraphics[width=\textwidth]{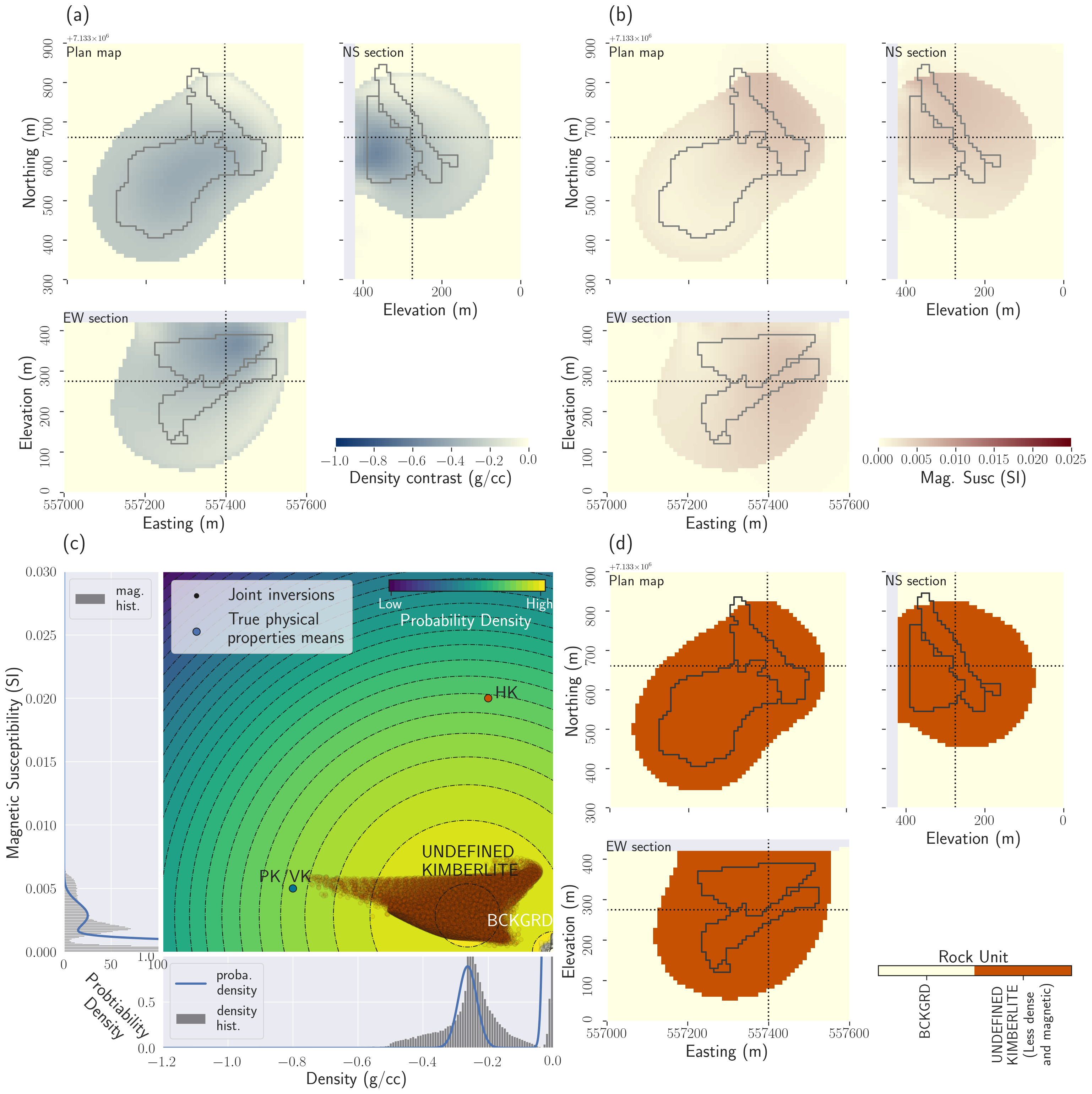}
    \caption{Results of the multi-physics PGI without providing the means of the physical properties for the kimberlite facies; a single kimberlite body is enough to meet the petrophysical requirements (the spread set by the covariances matrices) and reproduce the geophysical datasets; (a) Plan map, East-West and North-South cross-sections through the recovered density contrast model; (b) Plan map, East-West and North-South cross-sections through the magnetic susceptibility contrast model; (c) Cross-plot of the inverted models. The colour has been determined by the clustering obtained from this framework joint inversion process. In the background and side panels, we show the learned petrophysical GMM distribution; (d) Plan map, East-West and North-South cross-sections through the resulting quasi-geology model.}
    \label{fig:TKC_NoPrior_Synthetic.png}
\end{figure*}

\end{document}